\documentclass[onecolumn,doublespaced, 12pt]{IEEEtran}
\usepackage[english]{babel}
\usepackage{amsmath}
\usepackage{amssymb}
\usepackage{amsfonts}
\usepackage{graphicx}
\usepackage{psfrag}
\usepackage{epsfig}
\usepackage{bm}

\newcommand{\mb}{\mathbf}
\newcommand{\mr}{\mathrm}

\newtheorem{theorem}{Theorem}

\title{Large $n$ Analysis of Amplify-and-Forward MIMO Relay Channels with Correlated Rayleigh Fading}

\author{
{\it J\"{o}rg Wagner, Boris Rankov, and Armin Wittneben }\\
Wireless Communications Group, ETH Zurich\\
Sternwartstrasse 7, 8092 Zurich, Switzerland\\
\{jwagner, rankov, wittneben\}@nari.ee.ethz.ch}

\begin{document}
\maketitle

\begin{abstract}
In this correspondence the cumulants of the mutual information of
the flat Rayleigh fading amplify-and-forward MIMO relay channel
without direct link between source and destination are derived in
the large array limit. The analysis is based on the \emph{replica
trick} and covers both spatially independent and correlated fading
in the first and the second hop, while beamforming at all
terminals is restricted to deterministic weight matrices.
Expressions for mean and variance of the mutual information are
obtained. Their parameters are determined by a nonlinear equation
system. All higher cumulants are shown to vanish as the number of
antennas $n$ goes to infinity. In conclusion the distribution of
the mutual information $I$ becomes Gaussian in the large $n$ limit
and is completely characterized by the expressions obtained for
mean and variance of $I$. Comparisons with simulation results show
that the asymptotic results serve as excellent approximations for
systems with only few antennas at each node. The derivation of the
results follows the technique formalized by Moustakas et al. in
\cite{moustakas1}. Although the evaluations are more involved for
the MIMO relay channel compared to point-to-point MIMO channels,
the structure of the results is surprisingly simple again. In
particular an elegant formula for the mean of the mutual
information is obtained, i.e., the ergodic capacity of the two-hop
amplify-and-forward MIMO relay channel without direct link.
\end{abstract}


\section{Introduction}
\label{sec:introduction} Cooperative relaying has obtained major
attention in the wireless communications community in recent years
due to its various potentials regarding the enhancement of
diversity, achievable rates and range. An important milestone
within the wide scope of this field is the understanding of the
fundamental limits of the \emph{MIMO relay channel}. Such a
channel consists of a source, relay and destination terminal, each
equipped with multiple antennas.

Generally, there are different ways of including relays in the
transmission between a source and a destination terminal. Most
commonly relays are introduced to either decode the noisy signal
from the source or another relay, to re-encoded the signal and to
transmit it to another relay (\emph{multi-hop}) or the destination
terminal (\emph{two-hop}). Or the relay simply forwards a linearly
modified version of the noisy signal. These relaying strategies
are referred to as decode-and forward (DF) and amplify-and-forward
(AF), respectively. Currently, the simple AF approach seems to be
promising in many practical applications, e.g., since it is power
efficient, does not introduce any decoding delay and achieves full
diversity. Another approach is the so called compress-and-forward
strategy (CF), which quantizes the received signal and re-encodes
the resulting samples efficiently.

We briefly give an overview over important contributions to the
field of cooperative communications and relaying. The capability
of relays to provide diversity for combating multipath fading has
been studied in \cite{laneman04CoopDiversity},
\cite{lanemanTransIT03} and \cite{sendonaris03UserCoop}. In
\cite{WittnebenIST2003} the potential of spatial multiplexing gain
enhancement in correlated fading channels by means of relays has
been demonstrated. Tight upper and lower bounds on the capacity of
the fading relay channel are provided in \cite{KramerTIT2005,
GastparISIT2002, MadsenVTC2002, MadsenTIT2005}, and
\cite{KhojastepourCISS2003}. Furthermore, in \cite{nabarOyman03}
the capacity has been shown to scale like $N\log K$ for the fading
MIMO relay channel , where $N$ is the number of source and
destination antennas and $K$ is the number of relays.

In this paper we focus on the two-hop amplify-and-forward MIMO
relay channel with either i.i.d. or correlated Rayleigh fading
channel matrices. Our quantities of interest are the cumulant
moments of the mutual information of this channel. Of particular
importance in this context are its mean and variance. While the
mean completely determines the long term achievable rate in a fast
fading communication channel, the variance is crucial for the
characterization of the outage capacity of a channel, which is
commonly the quantity of interest in slow fading channels. Seeking
for closed form expressions of cumulant moments of the mutual
information in MIMO systems usually is a hopeless task. For the
conventional point-to-point MIMO channel it therefore turned out
to be useful to defer the analysis to the regime of large antenna
numbers. For the i.i.d. Rayleigh fading MIMO channel closed form
expressions were obtained in \cite{verdu} and \cite{mimomean1}.
For correlated fading at either transmitter or receiver side the
mean was derived \cite{mimomean3}, and \cite{mimomean4} finally
provided the mean for the case of MIMO interference. All these
results are obtained via the deterministic asymptotic eigenvalue
spectra of the respective matrices appearing in the capacity $\log
\det$-formula.

Higher moments were also considered, e.g., in \cite{mimostats1},
\cite{mimostats2} and \cite{moustakas1}, where the distribution in
the large antenna limit was identified to be Gaussian. Generally,
these large array results turned out to be very tight
approximations of the respective quantities in finite dimensional
systems. For amplify-and-forward MIMO relay channels only little
progress has been achieved so far even in the large array limit.
The mean mutual information of Rayleigh fading amplify-and-forward
MIMO relay channels in the large array limit has been studied in
\cite{venjamin} for the special case of a forwarding matrix
proportional to the identity matrix and uncorrelated channel
matrices. In this paper a fourth order equation for the Stieltjes
transform of the corresponding asymptotic eigenvalue spectrum is
found, which allows for a numerical evaluation of the mean mutual
information. Since even for this special case no analytic solution
is possible, the classical approach of evaluating the mean mutual
information via its asymptotic eigenvalue spectrum does not seem
to be promising for the AF MIMO relay channels.

The key tool enabling the evaluation of the cumulant moments of
the mutual information in the large array limit in this paper is
the so called replica method. It was introduced by Edwards and
Anderson in \cite{physics} and has its origins in physics where it
is applied to large random systems, as they arise, e.g., in
statistical mechanics. In the context of channel capacity it was
applied by Tanaka in \cite{capacity} for the first time. Moustakas
et al. \cite{moustakas1} finally used a framework utilizing the
replica trick developed in \cite{framework} to evaluate the
cumulant moments of the mutual information of the Rayleigh fading
MIMO channel in the presence of correlated interference. The paper
\cite{moustakas1} is formulated in a very explicatory way and this
correspondence goes very much along the lines of this reference.
Though not being proven in a rigorous way yet, the replica method
is a particularly attractive tool when dealing with functions of
large random matrices, since it allows for the evaluation of
arbitrary moments. Free probability theory, e.g., only allows for
the evaluation of the mean, e.g., \cite{rmtbook}. There are also
some large array results by M\"{u}ller that are of importance for
amplify-and-forward relay channels. He applied free probability
theory to concatenated vector fading channels in \cite{mueller1}
(two hops) and \cite{mueller2} (infinitely many hops), which can
be considered as multi-hop MIMO channels with noiseless relays.
The contributions of this paper are summarized as follows:
\begin{itemize}
    \item In the large array limit we derive mean and variance of the mutual information of the two-hop MIMO
            AF relay channel without direct link where the channel matrices are modelled as Kronecker correlated Rayleigh
            fading channels while the precoding matrix at the source and also the forwarding matrix at the relay
            are deterministic and constant over time. The obtained expression depends on coefficients that are
            determined by a system of six nonlinear equations.
    \item We show that all higher cumulant moments are ${\cal O}(n^{-1})$ or smaller
            and thus vanish as $n$ grows large. Accordingly, we conclude that the mutual information is Gaussian
            distributed with mean and variance given by our derived expressions in the large $n$ limit.
    \item Considering that not all doubts about the replica method are dispelled yet, we verify the obtained
            expressions by means of computer simulations and thus confirm that the replica method indeed works
            out in our problem.
\end{itemize}

\section{The Channel and its Mutual Information}

\begin{figure}[e]
        \centering \psfrag{lambda11}{$\lambda_1^{(1)}$}
        \psfrag{H1}{$\mb{H}_1$}
        \psfrag{G}{$\mb{F}_\mr{s}$}
         \psfrag{H2}{$\mb{H}_2$}
          \psfrag{F}{$\mb{F}_\mr{r}$}
           \psfrag{s}{$\mb{s}$}
            \psfrag{y}{$\mb{y}$}
            \psfrag{d}{$\mb{d}$}
             \psfrag{nr}{$\mb{n}_\mathrm{r}$}
             \psfrag{nd}{$\mb{n}_\mathrm{d}$}
             \psfrag{a}{{\tiny $n_\mathrm{s}$}}
             \psfrag{b}{{\tiny $n_\mathrm{r}$}}
             \psfrag{c}{{\tiny $n_\mathrm{r}$}}
             \psfrag{d}{{\tiny $n_\mathrm{d}$}}

        \includegraphics[width=\textwidth]{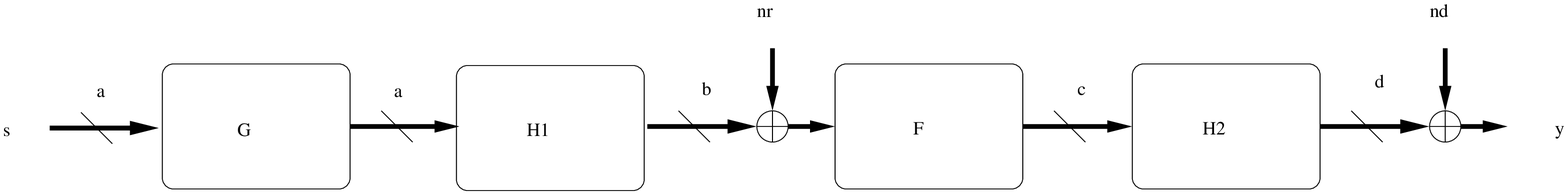}
        \caption{Block diagram of the channel.}
        \label{fig: block1}
    \end{figure}
\label{sec:model} The two-hop MIMO amplify-and-forward relay
channel under consideration is defined as follows. Three terminals
are equipped with $n_\mr{s}$ (source), $n_\mr{r}$ (relay), and
$n_\mr{d}$ (destination) antennas, respectively. We allow for
communication from source to relay and from relay to destination.
Particularly, we do not allot a direct communication link between
source and destination. Both the uplink (first hop from source to
relay) and the downlink (second hop from relay to destination) are
modelled as frequency-flat, i.e., the transmit symbol duration is
much longer than the delay spread of up- and downlink. We denote
the channel matrix of the uplink by $\mb{H}_1\in
\mathbb{C}^{n_\mr{r}\times n_\mr{s}}$, the one of the downlink by
$\mb{H}_2\in \mathbb{C}^{n_\mr{d}\times n_\mr{r}}$. Furthermore,
we assume that the relays process the received signals linearly.
The matrix performing this linear mapping is denoted
$\mb{F}_\mr{r}\in \mathbb{C}^{n_\mr{r}\times n_\mr{r}}$ and called
the ``forwarding matrix'' in the following. With $\mb{s}$ the
transmit symbol vector, a precoding matrix
$\mb{F}_\mr{s}\in\mathbb{C}^{n_\mr{s}\times n_\mr{s} }$ and
$\mb{n}_\mr{r}$ and $\mb{n}_\mr{d}$ the relay and destination
noise vectors respectively, the end-to-end input-output-relation
of this channel is then given by
\begin{equation}
\mb{y}=\mb{H}_2\mb{F}_\mr{r}\mb{H}_1\mb{F}_\mr{s}\mb{s}+\mb{H}_2\mb{F}_\mr{r}\mb{n}_\mr{r}+\mb{n}_\mr{d}.
\end{equation}
The system is represented in a block diagram in Fig. \ref{fig:
block1}.

The elements of the channel matrices $\mb{H}_1$ and $\mb{H}_2$
will be assumed to be zero mean circular symmetric complex
Gaussian (ZMCSCG) random variables with covariance matrices as
defined in the Kronecker model \cite{paulraj}:
\begin{eqnarray}
&& \mathsf{E}[\mathrm{vec}(\mb{H}_1)\mathrm{vec} (\mb{H}_1)^\mathrm{H}]=\mb{T}_\mr{s}^\mr{T}\otimes\mb{R}_\mr{r}, \textrm{ such that } \mr{Tr}\{\mb{T}_\mr{s}\}=n_\mr{s}\textrm{ and }\mr{Tr}\{\mb{R}_\mr{r}\}=n_\mr{r}\label{eq: kronecker1},\\
&& \mathsf{E}[\mathrm{vec}(\mb{H}_2)\mathrm{vec}
(\mb{H}_2)^\mathrm{H}]=\mb{T}_\mr{r}^\mr{T}\otimes\mb{R}_\mr{d},
\textrm{ such that } \mr{Tr}\{\mb{T}_\mr{s}\}=n_\mr{r}\textrm{ and
}\mr{Tr}\{\mb{R}_\mr{d}\}=n_\mr{d}\label{eq: kronecker2},
\end{eqnarray}
where $\mathrm{vec}(\mb{X})$ stacks $\mb{X}$ into a vector
columnwise, $\otimes$ denotes the Kronecker product while
$\mr{Tr}(\cdot)$ and $(\cdot )^\mr{T}$ denote the trace and
transposition operator, respectively. $\mb{T}_\mr{s}\in
\mathbb{C}^{n_\mr{s}\times n_\mr{s}}$, $\mb{R}_\mr{r}\in
\mathbb{C}^{n_\mr{r}\times n_\mr{r}}$, $\mb{T}_\mr{r}\in
\mathbb{C}^{n_\mr{r}\times n_\mr{r}}$ and $\mb{R}_\mr{d}\in
\mathbb{C}^{n_\mr{d}\times n_\mr{d}}$ are the (positive definite)
covariance matrices of the antenna arrays at the respective
terminals. These matrices are required to have full rank for the
analysis below. We remind the reader that matrices following
Gaussian distributions defined by covariance matrices as in
\eqref{eq: kronecker1} and \eqref{eq: kronecker2} can be generated
from a spatially white matrix $\mb{H}_w$ -- in our case through
the mappings
\begin{equation}
\mb{H}_1=\mb{R}_\mr{r}^\mr{\frac{1}{2}}\mb{H}_{w,1}\mb{T}_\mr{s}^\mr{\frac{1}{2}}
\end{equation}
and
\begin{equation}
\mb{H}_2=\mb{R}_\mr{d}^\mr{\frac{1}{2}}\mb{H}_{w,1}\mb{T}_\mr{r}^\mr{\frac{1}{2}}.
\end{equation}
The above described correlation model thus uses separable
correlations, which is a commonly accepted assumption for wireless
MIMO channels.

Since we will be confronted with products of covariance matrices
later on, we need to introduce the operator $(\cdot)^*$ for
quadratic matrices, which zeroizes the smaller of two matrices in
a product such that it adapts to the size of the other matrix, or
leaves it untouched if the matrix is the bigger one. Thus, we
ensure that products like $\mb{R}_\mr{r}^*\mb{R}_\mr{d}^*$ are
well defined. As long as two matrices $\mb{A}$ and $\mb{B}$ are
Toeplitz-like a product $\mb{A}^\mr{*}\mb{B}^\mr{*}$ always yields
the same result irrespective of the corner(s) used for zeroising
the smaller matrix.

We assume all channel matrix elements to be constant during a
certain interval and to change independently from interval to
interval (block fading). The input symbols are chosen to be i.i.d.
ZMCSCGs with variance $\rho$, i.e.,
$\mathsf{E}[\mb{s}\mb{s}^\mr{H}]=\rho/n_\mr{s}\mb{I}_{n_\mr{s}}$,
the additive noise at relay and destination is assumed to be white
in both space and time and is modelled as ZMCSCG with unit
variance, i.e.,
$\mathsf{E}[\mb{n}_\mr{r}\mb{n}_\mr{r}^\mr{H}]=\mb{I}_\mr{r}$ and
$\mathsf{E}[\mb{n}_\mr{d}\mb{n}_\mr{d}^\mr{H}]=\mb{I}_{n_\mr{d}}$.

The assumptions on the channel state information (CSI) are as
follows: The destination perfectly knows the instantaneous channel
matrices $\mb{H}_1$ and $\mb{H}_2$ as well as $\mb{F}_\mr{s}$ and
$\mb{F}_\mr{s}$. The source and the relay only know the second
order statistics of $\mb{H}_1$ and $\mb{H}_2$, i.e., the
corresponding covariance matrices. In particular this implies,
that the forwarding matrix can only depend on the covariance
matrices of $\mb{H}_1$ and $\mb{H}_2$, but not on the
instantaneous channel realizations. Its elements thus are
deterministic and remain constant over time. Our analysis could
only capture time-varying forwarding matrices that are Gaussian.
However, forwarding matrices chosen based on the current channel
realization would not be Gaussian in general. It will be useful to
decompose the forwarding matrix into a scaling factor
$\sqrt{\alpha/n_\mr{r}}$ and a matrix $\tilde{\mb{F}}_\mr{r}$
fulfilling
$\mathrm{Tr}\{\tilde{\mb{F}}_\mr{r}\tilde{\mb{F}}_\mr{r}^\mr{H}\}=n_\mr{r}$.
We will denote $\alpha$ as the power gain of the forwarding
matrix.

\begin{figure}[e]
        \centering \psfrag{lambda11}{$\lambda_1^{(1)}$}

           \psfrag{s}{$\mb{s}$}
            \psfrag{y}{$\mb{H}_1$}
            \psfrag{d}{$\mb{d}$}
             \psfrag{nr}{{\tiny$\mb{n}_\mathrm{r}$}}
             \psfrag{nd}{{\tiny$\mb{n}_\mathrm{d}$}}
             \psfrag{a}{{\tiny $n_\mathrm{s}$}}
             \psfrag{b}{{\tiny$n_\mathrm{r}$}}
             \psfrag{c}{{\tiny$n_\mathrm{d}$}}
             \psfrag{d}{$\mb{F}_\mathrm{s}$}
             \psfrag{e}{$\mb{T}_\mathrm{s}^\frac{1}{2}$}
             \psfrag{f}{$\mb{H}_{w,1}$}
             \psfrag{g}{$\mb{R}_\mathrm{r}^\frac{1}{2}$}
             \psfrag{h}{$\tilde{\mb{F}}_\mathrm{r}$}
              \psfrag{i}{$\mb{T}_\mathrm{r}^\frac{1}{2}$}
              \psfrag{j}{$\mb{H}_{w,2}$}
            \psfrag{k}{$\mb{R}_\mathrm{d}^\frac{1}{2}$}
             \psfrag{x}{$\tilde{\mb{H}}_{1}$}
              \psfrag{yy}{$\mb{y}$}
            \psfrag{z}{$\tilde{\mb{H}}_{2}$}
             \psfrag{zz}{$\mb{H}_2$}
              \psfrag{al}{$\frac{\alpha}{n_\mr{r}}$}

        \includegraphics[width=\textwidth]{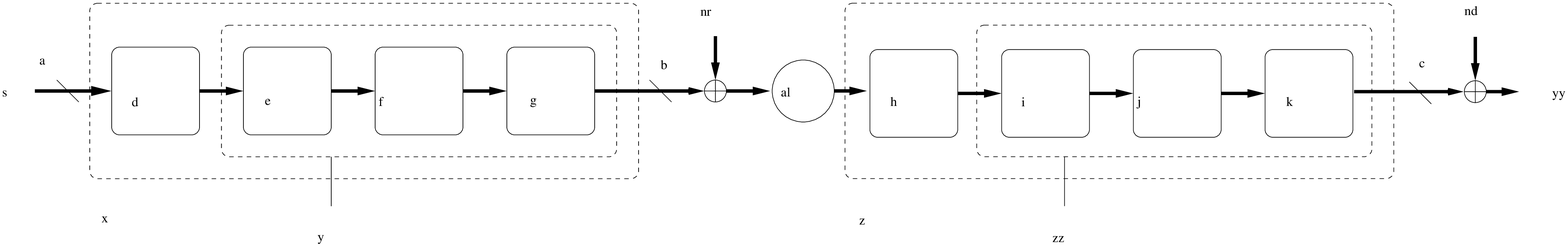}
        \caption{Extended block diagram of the channel.}
        \label{fig: block2}
    \end{figure}

With $\mathrm{Tr}\{\mb{F}_\mr{s}\mb{F}_\mr{s}^\mr{H}\}=n_\mr{s}$
the mutual information\footnote{In this chapter we pass on the
common pre-log factor $1/2$, which accounts for the use of two
time slots necessary in half-duplex relay protocols.} conditioned
on $\mb{H}_1$ and $\mb{H}_2$ in nats per channel use can be
written as
\begin{equation}\label{eq: mutualinformation}
I(\mb{s};\mb{y})=\ln\left(\frac{\det\left(\mb{I}_{n_\mr{d}}+\frac{\alpha}{n_\mr{r}}\mb{H}_2\tilde{\mb{F}}_\mr{r}\tilde{\mb{F}}_\mr{r}^\mr{H}\mb{H}_2^\mr{H}+\frac{\rho\cdot\alpha}{n_\mr{s}
n_\mr{r}}\mb{H}_2\tilde{\mb{F}}_\mr{r}\mb{H}_1\mb{F}_\mr{s}\mb{F}_\mr{s}^\mr{H}
\mb{H}_1^\mr{H}\tilde{\mb{F}}_\mr{r}^\mr{H}\mb{H}_2^\mr{H}\right)}{\det\left(\mb{I}_{n_\mr{d}}+\frac{\alpha}{n_\mr{r}}\mb{H}_2\tilde{\mb{F}}_\mr{r}\tilde{\mb{F}}_\mr{r}^\mr{H}\mb{H}_2^\mr{H}\right)}\right),
\end{equation}
where
\begin{equation}
\mb{I}_{n_\mr{d}}+\frac{\alpha}{n_\mr{r}}\mb{H}_2\tilde{\mb{F}}_\mr{r}\tilde{\mb{F}}_\mr{r}^\mr{H}\mb{H}_2^\mr{H}
\end{equation}
corresponds to the overall noise covariance matrix at destination
and \begin{equation} \frac{\rho\cdot\alpha}{n_\mr{s}
n_\mr{r}}\mb{H}_2\tilde{\mb{F}}_\mr{r}\mb{H}_1\mb{F}_\mr{s}\mb{F}_\mr{s}^\mr{H}
\mb{H}_1^\mr{H}\tilde{\mb{F}}_\mr{r}^\mr{H}\mb{H}_2^\mr{H}
\end{equation} corresponds to the signal plus noise covariance matrix at the
destination. Since the forwarding matrix does not depend on the
instantaneous channel realizations by assumption, it can be
incorporated into $\mb{T}_\mr{r}$ according to
\begin{equation}
\tilde{\mb{T}}_\mr{r}\triangleq
\tilde{\mb{F}}_\mr{r}\mb{T}_\mr{r}\tilde{\mb{F}}_\mr{r}^\mr{H}.
\end{equation}
Similarly $\mb{F}_\mr{s}$ can be incorporated into $\mb{T}_\mr{s}$
according to
\begin{equation}
\tilde{\mb{T}}_\mr{s}\triangleq
\mb{F}_\mr{s}\mb{T}_\mr{s}\mb{F}_\mr{s}^\mr{H}.
\end{equation}
Refer to the extended block diagram in Fig. \ref{fig: block2} for
an illustration.

In terms of the respective equivalent channels
$\tilde{\mb{H}}_1\triangleq
\mb{R}_\mr{r}^\frac{1}{2}\mb{H}_{w,1}\tilde{\mb{T}}_\mr{s}^\frac{1}{2}$
and $\tilde{\mb{H}}_2\triangleq
\mb{R}_\mr{d}^\frac{1}{2}\mb{H}_{w,2}\tilde{\mb{T}}_\mr{r}^\frac{1}{2}$
\eqref{eq: mutualinformation} can be rewritten as
\begin{equation}\label{eq: mutualinformationeq}
I(\mb{s};\mb{y})=\ln\left(\frac{\det\left(\mb{I}_{n_\mr{d}}+\frac{\alpha}{n_\mr{r}}\tilde{\mb{H}}_2\tilde{\mb{H}}_2^\mr{H}+\frac{\rho\cdot\alpha}{n_\mr{s}
n_\mr{r}}\tilde{\mb{H}}_2\tilde{\mb{H}}_1
\tilde{\mb{H}}_1^\mr{H}\tilde{\mb{H}}_2^\mr{H}\right)}{\det\left(\mb{I}\mr{d}+\frac{\alpha}{n_\mr{r}}\tilde{\mb{H}}_2\tilde{\mb{H}}_2^\mr{H}\right)}\right).
\end{equation}
In the subsequent sections we will work with \eqref{eq:
mutualinformationeq} and will drop the tildes again for the sake
of clarity.

Due to the randomness in $\mb{H}_1$ and $\mb{H}_2$ also $I$ is a
random variable. The theorem stated in the following section will
fully characterize the distribution of $I$ in the limit of large
antenna numbers.

\section{Results}

We formulate our results in the subsequent theorem. Whenever we
use the notation ${\cal O}(f(n))$ in the following we assume that
$n_\mr{s}$, $n_\mr{r}$ and $n_\mr{d}$ grow to infinity with all
ratios among them fixed.

\label{sec:results}
\begin{theorem}
For the mutual information $I$ as defined in \eqref{eq:
mutualinformationeq}

\begin{itemize}
\item the mean is ${\cal O}(n)$ and given by
\begin{eqnarray}
\mathsf{E}[I]&=&\ln\left(\det\left(\mb{I}_{n_\mr{s}}+\rho
s_1\mb{T}_\mr{s}\right)\right)+\ln\left(\det\left(\mb{I}_{n_\mr{r}}+\alpha
s_2\mb{T}_\mr{r}\right)\right)\nonumber\\
&&\hspace{.5cm}-\ln\left(\det\left(\mb{I}_{n_\mr{r}}+\alpha
s_3\mb{T}_\mr{r}\right)\right)-\ln\left(\det\left(\mb{I}_{n_\mr{d}}+t_3\mb{R}_\mr{d}\right)\right)\nonumber\\
&&\hspace{1.25cm}+\ln\left(\det\left(\mb{I}_{\max(n_\mr{r},n_\mr{d})}+t_2\mb{R}_\mr{d}^*+t_1t_2\mb{R}_\mr{r}^*\mb{R}_\mr{d}^*\right)\right)\nonumber\\
&&\hspace{2cm}-\left(n_\mr{s} s_1t_1+n_\mr{r} s_2t_2-n_\mr{r}
s_3t_3\right)+{\cal O}(n^{-1})\label{eq: mean}
\end{eqnarray}
with
\begin{eqnarray}
n_\mr{s} t_1&=&\mathrm{Tr}\left\{\rho \mb{T}_\mr{s}\left[\mb{I}_{n_\mr{s}}+\rho s_1 \mb{T}_\mr{s}\right]^{-1}\right\}\\
n_\mr{r} t_2&=&\mathrm{Tr}\left\{\alpha \mb{T}_\mr{r}\left[\mb{I}_{n_\mr{r}}+\alpha s_2 \mb{T}_\mr{r}\right]^{-1}\right\}\\
n_\mr{r} t_3&=&\mathrm{Tr}\left\{\alpha
\mb{T}_\mr{r}\left[\mb{I}_{n_\mr{r}}+\alpha s_3
\mb{T}_\mr{r}\right]^{-1}\right\}\\
n_\mr{s} s_1&=&\mathrm{Tr}\left\{t_2\mb{R}_\mr{r}^*\mb{R}_\mr{d}^*\left[\mb{I}_{n_\mr{r}}+t_1t_2\mb{R}_\mr{r}^*\mb{R}_\mr{d}^*+t_2\mb{R}_\mr{d}^*\right]^{-1}\right\}\\
n_\mr{r} s_2&=&\mathrm{Tr}\left\{(\mb{R}_\mr{r}+t_2\mb{R}_\mr{r}^*\mb{R}_\mr{d}^*)\left[\mb{I}_{n_\mr{r}}+t_1t_2\mb{R}_\mr{r}^*\mb{R}_\mr{d}^*+t_2\mb{R}_\mr{d}^*\right]^{-1}\right\}\\
n_\mr{r}
s_3&=&\mathrm{Tr}\left\{\mb{R}_\mr{d}\left[\mb{I}_{n_\mr{r}}+t_3
\mb{R}_\mr{d}\right]^{-1}\right\},
\end{eqnarray}

\item the variance is ${\cal O}(1)$ and given by
\begin{equation}
\mathsf{Var}[I]=-\ln(\det(\mb{V}_1))-\ln(\det(\mb{V}_2))+2\ln(\det(\mb{V}_3))+{\cal
O}(n^{-2})
\end{equation}
with
\begin{equation}\label{eq: V1}
\mb{V}_1=
\begin{pmatrix}
  v_1^{(1)} & 1 & 0 & 0 & 0 & 0 & 0 & 0 & 0 & 0 \\
  1 & v_{12}^{(1)} & 0 & v_3^{(1)} & 0 & v_8^{(1)} & 0 & v_8^{(1)} & 0 & v_4^{(1)} \\
  0 & 0 & v_2^{(1)} & 1 & v_2^{(1)} & 0 & v_2^{(1)} & 0 & v_2^{(1)} & 0 \\
  0 & v_3^{(1)} & 1 & v_5^{(1)} & 0 & v_9^{(1)} & 0 & v_9^{(1)} & 0 & v_6^{(1)} \\
  0 & 0 & v_2^{(1)} & 0 & v_2^{(1)} & 1 & v_2^{(1)} & 0 & v_2^{(1)} & 0 \\
  0 & v_8^{(1)} & 0 & v_9^{(1)} & 1 & v_6^{(1)} & 0 & v_{10}^{(1)} & 0 & v_{11}^{(1)} \\
  0 & 0 & v_2^{(1)} & 0 & v_2^{(1)} & 0 & v_2^{(1)} & 1 & v_2^{(1)} & 0 \\
  0 & v_8^{(1)} & 0 & v_9^{(1)} & 0 & v_{10}^{(1)} & 1 & v_6^{(1)} & 0 & v_{11}^{(1)}\\
  0 & 0 & v_2^{(1)} & 0 & v_2^{(1)} & 0 & v_2^{(1)} & 0 & v_2^{(1)} & 1 \\
  0 & v_4^{(1)} & 0 & v_6^{(1)} & 0 & v_{11}^{(1)} & 0 & v_{11}^{(1)} & 1 & v_7^{(1)}
\end{pmatrix}
\end{equation}
\begin{equation}\label{eq: V2}
\mb{V}_2=\begin{pmatrix}
  v_1^{(2)} & 1 \\
  1 & v_2^{(2)}
\end{pmatrix}
\end{equation}
\begin{equation}\label{eq: V3}
\mb{V}_3=\begin{pmatrix}
  1 & -v_1^{(3)} & 0 & v_1^{(3)} \\
  -v_4^{(3)} & 1 & v_2^{(3)} & 0 \\
 0 & v_1^{(3)} & -1 & -v_1^{(3)} \\
  v_2^{(3)} & 0 & -v_3^{(3)} & -1
\end{pmatrix}
\end{equation}
and
\begin{eqnarray}
v_1^{(1)}&=&\frac{1}{n_\mr{s}^2}\mathrm{Tr}\left\{\left(\rho \mb{T}_\mr{s}\left[\mb{I}+\rho s_1\mb{T}_\mr{s}\right]^{-1}\right)^2\right\}\\
v_2^{(1)}&=&\frac{1}{n_\mr{r}^2}\mathrm{Tr}\left\{\left(\alpha \mb{T}_\mr{r}\left[\mb{I}+\alpha s_2\mb{T}_\mr{r}\right]^{-1}\right)^2\right\}\\
v_3^{(1)}&=&-\mathrm{Tr}\left\{\mb{R}_\mr{r}^*\mb{R}_\mr{d}^*\left(\left[\mb{I}+t_2 \mb{R}_\mr{d}^*\right]\left[\mb{I}+t_2\mb{R}_\mr{d}^*+t_1t_2\mb{R}_\mr{r}^*\mb{R}_\mr{d}^*\right]^{-1}\right)^2\right\}\\
v_4^{(1)}&=&-\mathrm{Tr}\left\{\mb{R}_\mr{r}\left(t_2\mb{R}_\mr{d}^*\left[\mb{I}+t_2\mb{R}_\mr{d}^*+t_1t_2\mb{R}_\mr{r}^*\mb{R}_\mr{d}^*\right]^{-1}\right)^2\right\}\\
v_{12}^{(1)}&=&\mathrm{Tr}\left\{\left(t_2\mb{R}_\mr{r}^*\mb{R}_\mr{d}^*\left[\mb{I}+t_2\mb{R}_\mr{d}^*+t_1t_2\mb{R}_\mr{r}^*\mb{R}_\mr{d}^*\right]^{-1}\right)^2\right\}\\
v_5^{(1)}&=&\mathrm{Tr}\left\{\left(t_1\mb{R}_\mr{r}^*\mb{R}_\mr{d}^*\left[\mb{I}+t_2\mb{R}_\mr{d}^*\right]\left[\mb{I}+t_2\mb{R}_\mr{d}^*+t_1t_2\mb{R}_\mr{r}^*\mb{R}_\mr{d}^*\right]^{-1}\right)^2\right\}\\
v_6^{(1)}&=&\mathrm{Tr}\left\{\left(t_1t_2\mb{R}_\mr{r}^*\mb{R}_\mr{d}^{*2}\left[\mb{I}+t_2\mb{R}_\mr{d}^*+t_1t_2\mb{R}_\mr{r}^*\mb{R}_\mr{d}^*\right]^{-1}\right)^2\right\}\\
v_7^{(1)}&=&\mathrm{Tr}\left\{\left(\left[\mb{R}_\mr{d}^*+t_1t_2\mb{R}_\mr{r}^*\mb{R}_\mr{d}^{*2}\right]\left[\mb{I}+t_2\mb{R}_\mr{d}^*+t_1t_2\mb{R}_\mr{r}^*\mb{R}_\mr{d}^*\right]^{-1}\right)^2\right\}\\
v_8^{(1)}&=&\mathrm{Tr}\left\{t_2\mb{R}_\mr{r}^*\mb{R}_\mr{d}^*\mb{R}_\mr{d}^*\left[\mb{I}+t_2 \mb{R}_\mr{d}^*\right]\left(\left[\mb{I}+t_2\mb{R}_\mr{d}^*+t_1t_2\mb{R}_\mr{r}^*\mb{R}_\mr{d}^*\right]^{-1}\right)^2\right\}\\
v_9^{(1)}&=&-\mathrm{Tr}\left\{t_2\mb{R}_\mr{d}^*\left[\mb{I}+t_2\mb{R}_\mr{d}^*\right]\left(t_1\mb{R}_\mr{r}^*\mb{R}_\mr{d}^*\left[\mb{I}+t_2\mb{R}_\mr{d}^*+t_1t_2\mb{R}_\mr{r}^*\mb{R}_\mr{d}^*\right]^{-1}\right)^2\right\}\\
v_{10}^{(1)}&=&\mathrm{Tr}\bigg\{t_1\mb{R}_\mr{r}^*\mb{R}_\mr{d}^{*2}\left[\mb{I}+t_2\mb{R}_\mr{d}^*\right]\left[\mb{I}+t_1t_2\mb{R}_\mr{r}^*\mb{R}_\mr{d}^*\right]\nonumber\\
&&\hspace{1cm}\times\left(\left[\mb{I}+t_2\mb{R}_\mr{d}^*+t_1t_2\mb{R}_\mr{r}^*\mb{R}_\mr{d}^*\right]^{-1}\right)^2\bigg\}\\
v_{11}^{(1)}&=&-\mathrm{Tr}\left\{t_1t_2\mb{R}_\mr{r}^*\mb{R}_\mr{d}^{*2}\left[\mb{I}+t_1t_2\mb{R}_\mr{r}^*\mb{R}_\mr{d}^*\right]\left(\left[\mb{I}+t_2\mb{R}_\mr{d}^*+t_1t_2\mb{R}_\mr{r}^*\mb{R}_\mr{d}^*\right]^{-1}\right)^2\right\}\\
v_1^{(2)}&=&\frac{1}{n_\mr{r}^2}\mathrm{Tr}\left\{\left(\alpha \mb{T}_\mr{r}\left[\mb{I}+\alpha s_3\mb{T}_\mr{r}\right]^{-1}\right)^2\right\}\\
v_2^{(2)}&=&\mathrm{Tr}\left\{\left(\mb{R}_\mr{d}\left[\mb{I}+ t_3\mb{R}_\mr{d}\right]^{-1}\right)^2\right\}\\
v_1^{(3)}&=&\frac{1}{n_\mr{r}}\mathrm{Tr}\left\{\alpha \mb{R}_\mr{d}^*\mb{T}_\mr{r}^*\left[\mb{I}+ t_3\mb{R}_\mr{d}^*\right]^{-1}\left[\mb{I}+ \alpha s_2\mb{T}_\mr{r}^*\right]^{-1}\right\}\\
v_2^{(3)}&=&\frac{1}{n_\mr{r}}\mathrm{Tr}\left\{t_1t_2\mb{R}_\mr{d}^*\mb{T}_\mr{r}^* \mb{R}_\mr{r}^*\mb{R}_\mr{d}^*\left[\mb{I}+t_2\mb{R}_\mr{d}^*+t_1t_2\mb{R}_\mr{r}^*\mb{R}_\mr{d}^*\right]^{-1}\left[\mb{I}+\alpha s_3 \mb{T}_{2}^*\right]^{-1}\right\}\\
v_3^{(3)}&=&\frac{1}{n_\mr{r}}\mathrm{Tr}\big\{\mb{R}_\mr{d}^*\mb{T}_\mr{r}^*\left[\mb{I}+t_1t_2 \mb{R}_\mr{r}^*\mb{R}_\mr{d}^*\right]\nonumber\\
&&\hspace{1cm}\times\left[\mb{I}+t_2\mb{R}_\mr{d}^*+t_1t_2\mb{R}_\mr{r}^*\mb{R}_\mr{d}^*\right]^{-1}\left[\mb{I}+\alpha s_3 \mb{T}_{2}^*\right]^{-1}\big\}\\
v_4^{(3)}&=&\frac{1}{n_\mr{r}}\mathrm{Tr}\big\{\mb{R}_\mr{d}^*\mb{T}_\mr{r}^*\left[t_1\mb{R}_\mr{r}^*+t_1t_2
\mb{R}_\mr{r}^*\mb{R}_\mr{d}^*\right]\nonumber\\
&&\hspace{1cm}\times\left[\mb{I}+t_2\mb{R}_\mr{d}^*+t_1t_2\mb{R}_\mr{r}^*\mb{R}_\mr{d}^*\right]^{-1}\left[\mb{I}+\alpha
s_3 \mb{T}_{2}^*\right]^{-1}\big\},
\end{eqnarray}

\item all higher cumulant moments (skewness, kurtosis, etc.) are
${\cal} O(n^{-1})$ or smaller and thus vanish, as $n$ grows large.
Consequently, the mutual information $I$ is Gaussian distributed
random variable in the large $n$ limit.

\end{itemize}
\end{theorem}

\section{Mathematical Tools}
In this section we briefly repeat the mathematical tools we will
use in the proof of the theorem. These are (cumulant) moment
generating functions, the replica method and saddle point
integration. At the same time we shall give a brief outline of the
proof, which we will provide in full detail in Section
\ref{sec:proof}.

\subsection{Generating Functions}

We define the \emph{moment generating function} of the mutual
information $I$ as follows:
\begin{equation}\label{eq: mom}
g_I(\nu)=\mathsf{E}\left[e^{-\nu I}\right].
\end{equation}
This definition differs from the standard definition in the sign
of the argument of the exponential function. The minus sign used
in the definition above will simplify our analysis later on.
Provided that the moment generating function exists in an interval
around $\nu=0$ we may expand \eqref{eq: mom} into a series in the
following way
\begin{equation}\label{eq: mom2}
g_I(\nu)=1-\nu\cdot\mathsf{E}[I]+\frac{\nu^2}{2}\cdot\mathsf{E}[I^2]-\frac{\nu^3}{6}\cdot\mathsf{E}[I^3]+\ldots.
\end{equation}
We
will also consider the \emph{cumulant generating function} of $I$,
which is defined as $\ln g_I(\nu)$. Expanded into a Taylor series
around zero it is given by
\begin{equation}\label{eq: cumgen}
\ln(g_I(\nu))=-\nu\cdot \mathsf{E}[\ln f(X)]+\frac{\nu^2}{2}
\cdot\mathsf{Var}[\ln
f(X)]+\sum_{p=3}^\infty\frac{(-\nu)^p}{p!}{\cal C}_p,
\end{equation}
with ${\cal C}_p$ the $p$th cumulant moment. Once we have found
this series, it is thus easy to extract mean and variance by a
simple comparison of coefficients. Furthermore, since a Gaussian
random variable has the unique property that only a finite number
of its cumulants are nonzero (more precisely its mean and
variance), we will be able to proof the asymptotic Gaussianity of
$I$ by showing that the cumulants ${\cal C}_p$ die out for $p>2$
in the large $n$ limit.

\subsection{Integral Identities}
\label{sec:complexintegrals}We will need some useful integral
identities in order to evaluate the moment generating function.
Before stating them we introduce a compact notation for products
of differentials arising when integration over elements of
matrices is performed. With $\imath=\sqrt{-1}$ as well as $\Re{Z}$
and $\Im{Z}$ the real and imaginary part of a complex variable
$Z$, we define the following integral measures, which are
completely identical with the notation used in \cite{moustakas1}:
\begin{eqnarray}
d_c\mb{X}&\triangleq &\frac{1}{2\pi}\prod_i\prod_j d\Re{X_{ij}}d
\Im{X_{ij}}, \textrm{ for $X_{ij}$ complex variables},\\
d_g\mb{X}&\triangleq &\frac{1}{2\pi}\prod_i\prod_j d X_{ij} d \overline{X}_{ij}, \textrm{ for $X_{ij}$ and $\overline{X}_{ij}$ Grassmann variables},\\
d\mu (\mb{X},\mb{Y})&\triangleq &\frac{1}{ 2\pi
\imath}\prod_i\prod_j d X_{ij}d Y_{ji}, \textrm{ for $X_{ij}$ and
$Y_{ij}$ complex variables}.
\end{eqnarray}
The defining properties of a Grassmann variable are listed in
Appendix \ref{sec: grassmann}. With this notation as well as
$\otimes$ the Kronecker product operator we specify the following
identities, which are all proven in \cite{moustakas1}:
\begin{itemize}
\item For $\mb{M}\in\mathbb{C}^{n\times n}$,
$\mb{N}\in\mathbb{C}^{\nu\times \nu}$ positive definite, and
$\mb{X},\mb{A},\mb{B}\in\mathbb{C}^{n\times \nu}$ we have
\begin{eqnarray}\label{eq: identity1}
&&\int\exp\left(-\frac{1}{2}\mathrm{Tr}\left(\mb{N}\mb{X}^\mathrm{H}\mb{M}\mb{X}+\mb{A}^\mathrm{H}\mb{X}-\mb{X}^\mathrm{H}\mb{B}\right)\right)
d_c\mb{X}\nonumber\\
&&\hspace{2cm}=\left(\det\left(\mb{N}\otimes
\mb{M}\right)\right)^{-1}\exp\left(-\frac{1}{2}\mathrm{Tr}\left(\mb{N}^{-1}\mb{A}^\mathrm{H}\mb{M}^{-1}\mb{B}\right)\right).
\end{eqnarray}

\item For $\mb{M}\in\mathbb{C}^{n\times n}$,
$\mb{N}\in\mathbb{C}^{\nu\times \nu}$ positive definite, and
$\overline{\mb{A}},\overline{\mb{X}}$ and $\mb{X}, \mb{B}$
$n\times \nu$ and $\nu\times n$ matrices, respectively, whose
entries are Grassmann variables,
 we have
\begin{eqnarray}\label{eq: identity2}
&&\int\exp\left(\mathrm{Tr}\left(\mb{N}\overline{\mb{X}}\mb{M}\mb{X}+\overline{\mb{A}}\mb{X}+\overline{\mb{X}}\mb{B}\right)\right)
d_g\mb{X}\nonumber\\&&\hspace{3cm}=\det\left(\mb{N}\otimes
\mb{M}\right)\exp\left(\mathrm{Tr}\left(\mb{N}^{-1}\overline{\mb{A}}\mb{M}^{-1}\mb{B}\right)\right).\end{eqnarray}

\item For $\mb{X},\mb{Y},\mb{A},\mb{B}\in\mathbb{C}^{\nu\times
\nu}$ we have
\begin{equation}\label{eq: identity3}
\int\exp\left(\mathrm{Tr}\left(\mb{X}\mb{Y}-\mb{X}\mb{A}-\mb{B}\mb{Y}\right)\right)
d\mu
(\mb{X},\mb{Y})=\exp\left(-\mathrm{Tr}\left(\mb{A}\mb{B}\right)\right).
\end{equation}
\end{itemize}
The application of these identities is known as the \emph{replica
trick}, which introduces multiple copies of the Gaussian
integration that arises when computing the expectation of
$\exp(-\nu I)$ over the elements of $\mb{H}_1$ and $\mb{H}_2$. We
emphasize that the machinery of repeatedly applying the above
identities in the evaluation of $g_I(\nu)$ (see Section \ref{sec:
step1}) requires $\nu$ to be a positive integer. In order to
extract the (cumulant) moments of $I$ from the respective
generating function we thus will need to assume that $g_I(\nu)$
can be analytically continued at least in the positive vicinity of
zero in the end. This assumption will be applied without being
proven anywhere in the literature yet. Nevertheless, all results
obtained based on this assumption -- including those derived below
-- show a perfect match with results obtained through computer
simulations.

\subsection{Saddle Point Integration}
\label{subsec: saddlepoint} For the final evaluation of the moment
generating function we will use the saddle point method. In its
simplest form it is an useful tool to solve integrals of the form
\begin{equation}
\lim_{n\rightarrow \infty}\int e^{-n\cdot
\Psi(\mb{}x_1,\ldots,x_k)}\cdot dx_1\cdots dx_k,
\end{equation}
where $\Psi(\cdot,\ldots,\cdot)$ is some function with well
defined Hessian at its global minimum. We will use it with a
slightly different expression in this paper. 
For the sake of clarity we will consider the univariate case in
this section. In the actual proof of the Theorem we will then deal
with integrals over multiple variables. Suppose we can rewrite the
moment generating function of $I$ in the form (as done in Section
\ref{sec: step1})
\begin{equation}
g_I(\nu)=\int e^{-f(x,\nu,n)} dx.
\end{equation}
If we expand $f(\cdot)$ into a Taylor series in $x$ around its
global minimum in $x_0$ we can write
\begin{equation}\label{eq: explanation}
g_I(\nu)=e^{-f(x_0,\nu,n)}\cdot\int
e^{-\frac{1}{2}f^{''}(x_0,\nu,n)(x-x_0)^2-\frac{1}{6}f^{'''}(x_0,\nu,n)(x-x_0)^3+\ldots}
\cdot dx,
\end{equation}
where the $(\cdot)^{'}$ operator denotes derivation for $x$. From
this expansion and our particular function $f(\cdot,\nu,n)$, which
will be multivariate indeed, it will possible to show that
\eqref{eq: explanation} evaluates to
\begin{equation}\label{eq: saddle}
g_I(\nu)=\exp\left(-\nu \cdot n \cdot \xi_1(x_0)+\nu^2 \cdot
\xi_2(x_0) \right)+\sum_{k\geq 3} \nu^k \cdot {\cal O}(n^{-1}),
\end{equation}
with $\xi_1(\cdot)$ and $\xi_2(\cdot)$ functions that we determine
in Section \ref{sec: step2}. The fact that
\begin{equation}\label{eq: momx}
\mathsf{E}\left[I^p\right]=(-1)^p\cdot\left(\frac{d^p}{d\nu^p}\hspace{.2cm}
g_I(\nu)\right)\Bigg|_{\nu=0},
\end{equation}
immediately reveals that the leading terms of mean and variance
are determined by $\xi_1$ and $\xi_2$, respectively. The ${\cal
O}(n^{-1})$ scaling of the residual terms is proven in Section
\ref{sec: step3}. Comparing $\ln g_I(\nu)$ to the right hand side
of \eqref{eq: cumgen} will reveal the higher order cumulants to be
at most ${\cal O}(n^{-1})$. Remember that we obtained \eqref{eq:
cumgen} as a Taylor expansion around $\nu=0$. We thus have
implicitly assumed that the limit $n\rightarrow \infty$ and
$\nu\rightarrow 0$ can be interchanged. This assumption is
noncritical and made without proof in this paper.

%

In the subsequent sections we will apply this procedure in a
multivariate framework. $f(\cdot, \nu,n)$ will then be a function
of multiple matrices with a appropriately defined integration
measures (cf. next subsection), which appear inside trace and
determinant operators. We will make a symmetry assumption called
the hypothesis of replica invariance, namely that all these
matrices are proportional to the identity matrix at the global
minimum of $f(\cdot,\nu,n)$. This assumption is justified in
\cite{sengupta}. Therefore no proof is provided in this paper.

We highlight that it is this saddle point method that makes the
following derivations a large $n$ approximation. If we had another
tool capable to solve the critical integral for finite $n$ the
whole procedure could also be applied to obtain nonasymptotic
results.

\section{Proof}
\label{sec:proof}

The equations in the proof\footnote{This proof is only rigorous in
the case that the analytic continuation of $g_I(\nu)$ to zero is
indeed possible. Proving this in turn is a current research topic
in mathematics.} are somewhat involved. In order to make the proof
clearly laid out and more compact we therefore omit the channel
covariance matrices and assume antenna arrays of size $n$ at each
terminal at first instance. Both covariance matrices and the
possibly different antenna numbers can be easily reintroduced at
the end of the proof. For the sake of clarity we structure the
proof into three parts corresponding to the subsections below.

\subsection{Applying the Replica Trick}
\label{sec: step1} We introduce the auxiliary variables
$\mb{X},\mb{Y},\mb{Z},\mb{W}_1,\mb{W}_2\in\mathbb{C}^{n\times
\nu}$ and $\overline{\mb{A}},\overline{\mb{B}},\mb{A}\mb{B}$ ($\nu
\times n$ and $n \times \nu$ Grassmann matrices) and evaluate the
moment generating function of $I$ by means of identities
\eqref{eq: identity1} - \eqref{eq: identity3} as follows
\begin{eqnarray}
g_I(\nu)&=&\mathsf{E}[e^{-\nu
I}]=\mathsf{E}\left[\left\{\frac{\det\left(\mb{I}+\frac{\alpha}{n}\mb{H}_2\mb{H}_2^{\mathrm{H}}+\frac{\rho\alpha}{n^2}\mb{H}_2\mb{H}_1\mb{H}_1^{\mathrm{H}}\mb{H}_2^{\mathrm{H}}\right)}{\det
\left(
\mb{I}+\frac{\alpha}{n}\mb{H}_2\mb{H}_2^{\mathrm{H}}\right)}\right\}^{-\nu}\right]\\
&=&\int \bigg[\int \exp \left(-\frac{1}{2}\mathrm{Tr}\left(\mb{X}^\mathrm{H}\mb{X}+\mb{Y}^\mathrm{H}\mb{Y}+\mb{Z}^\mathrm{H}\mb{Z}\right)\right)\nonumber\\
&&\times\exp \left(-\frac{1}{2}\mathrm{Tr}\left(\mb{Y}^\mathrm{H}\mb{H}_1^\mathrm{H}\mb{H}_2^\mathrm{H}\mb{X}-\mb{X}^\mathrm{H}\mb{H}_2\mb{H}_1\mb{Y}+\frac{\alpha}{n}\mb{Z}^\mathrm{H}\mb{H}_2^\mathrm{H}\mb{X}-\mb{X}^\mathrm{H}\mb{H}_2\mb{Z}\right)\right)\nonumber\\
&&\hspace{3cm}\times d_c\mb{X}d_c\mb{Y}d_c\mb{Z}\bigg]\nonumber\\
&&\times \left[\int  \exp
\left(\mathrm{Tr}\left(\overline{\mb{A}}\mb{A}+\overline{\mb{B}}\mb{B}+\frac{\alpha}{n}\overline{\mb{B}}\mb{H}_2^\mathrm{H}\mb{A}-\overline{\mb{A}}\mb{H}_2\mb{B}\right)\right)d_g\mb{A}d_g\mb{B}\right]
 \nonumber\\
 &&\hspace{3cm}\times\exp\left(-\mathrm{Tr}\left(\mb{H}_1^\mathrm{H}\mb{H}_1+\mb{H}_2^\mathrm{H}\mb{H}_2 \right)\right)  d\mb{H}_1d\mb{H}_2\label{eq: moment1}\\
&=&\int \left[\int\bigg[\int \exp \left(-\frac{1}{2}\mathrm{Tr}\left(\mb{X}^\mathrm{H}\mb{X}+\mb{Y}^\mathrm{H}\mb{Y}+\mb{Z}^\mathrm{H}\mb{Z}+\mb{W}_1^\mathrm{H}\mb{W}_1+\mb{W}_2^\mathrm{H}\mb{W}_2\right)\right)\right.\nonumber\\
&&\times\exp\left(-\frac{1}{2}\mathrm{Tr}\left(\frac{\rho}{n}\mb{Y}^\mathrm{H}\mb{H}_1^\mathrm{H}\mb{W}_2-\frac{\alpha}{n}\mb{W}_2^\mathrm{H}\mb{H}_2^\mathrm{H}\mb{X}+\mb{X}^\mathrm{H}\mb{H}_2\mb{W}_1+\mb{W}_1^\mathrm{H}\mb{H}_1\mb{Y}\right)\right)\nonumber\\
&&\hspace{3cm}\times d_c\mb{W}_1d_c\mb{W}_2\bigg]\nonumber\\
&&\left.\times\exp\left(-\frac{1}{2}\mathrm{Tr}\left(\frac{\alpha}{n}\mb{Z}^\mathrm{H}\mb{H}_2^\mathrm{H}\mb{X}-\mb{X}^\mathrm{H}\mb{H}_2\mb{Z}\right)\right)d_c\mb{X}d_c\mb{Y}d_c\mb{Z}\right]\nonumber\\
&&\times \left[\int
\exp\left(\mathrm{Tr}\left(\overline{\mb{A}}\mb{A}+\overline{\mb{B}}\mb{B}+\frac{\alpha}{n}\overline{\mb{B}}\mb{H}_2^\mathrm{H}\mb{A}-\overline{\mb{A}}\mb{H}_2\mb{B}\right)\right)d_g\mb{A}d_g\mb{B}\right]\nonumber\\
&&\times \exp\left(-\mathrm{Tr}\left(\mb{H}_1^\mathrm{H}\mb{H}_1+\mb{H}_2^\mathrm{H}\mb{H}_2 \right)\right)  d\mb{H}_1d\mb{H}_2\label{eq: moment1b}\\
&=&\int  \exp \left(-\frac{1}{2}\mathrm{Tr}\left(\mb{X}^\mathrm{H}\mb{X}+\mb{Y}^\mathrm{H}\mb{Y}+\mb{Z}^\mathrm{H}\mb{Z}+\mb{W}_1^\mathrm{H}\mb{W}_1+\mb{W}_2^\mathrm{H}\mb{W}_2-2\overline{\mb{A}}\mb{A}-2\overline{\mb{B}}\mb{B}\right)\right)\nonumber\\
&&\times\exp\left(-\frac{1}{4}\mathrm{Tr}\left(-\frac{
\rho}{n}\mb{Y}^\mathrm{H}\mb{Y}\mb{W}_1^\mathrm{H}\mb{W}_2+\frac{\alpha
}{n}\mb{X}^\mathrm{H}\mb{X}\mb{W}_2^\mathrm{H}\mb{W}_1-\frac{\alpha}{n}\mb{X}^\mathrm{H}\mb{X}\mb{Z}^\mathrm{H}\mb{W}_1\right)\right)\nonumber\\
&&\times\exp\left(-\frac{1}{4}\mathrm{Tr}\left(2\frac{\alpha}{n}\overline{\mb{B}}\mb{W}_1\mb{X}^\mathrm{H}\mb{A}-\frac{\alpha}{n}\mb{X}^\mathrm{H}\mb{X}\mb{W}_2^\mathrm{H}\mb{Z}+\frac{\alpha}{n}\mb{X}^\mathrm{H}\mb{X}\mb{Z}^\mathrm{H}\mb{Z}\right)\right)\nonumber\\
&&\times\exp\left(-\frac{1}{4}\mathrm{Tr}\left(-2\frac{\alpha}{n}\overline{\mb{B}}\mb{Z}\mb{X}^\mathrm{H}\mb{A}+2\frac{\alpha}{n}\overline{\mb{A}}\mb{X}\mb{W}_2^\mathrm{H}\mb{B}-2\frac{\alpha}{n}\overline{\mb{A}}\mb{X}\mb{Z}^\mathrm{H}\mb{B}-4\frac{\alpha}{n}\overline{\mb{B}}\mb{B}\overline{\mb{A}}\mb{A}
\right)\right)\nonumber\\&&\hspace{3cm}\times d_c\mb{W}_1d_c\mb{W}_2d_c\mb{X}d_c\mb{Y}d_c\mb{Z}d_g\mb{A}d_g\mb{B}\label{eq: moment2}\\
&=&\int  \exp \left(-\frac{1}{2}\mathrm{Tr}\left(\mb{X}^\mathrm{H}\mb{X}+\mb{Y}^\mathrm{H}\mb{Y}+\mb{Z}^\mathrm{H}\mb{Z}+\mb{W}_1^\mathrm{H}\mb{W}_1+\mb{W}_2^\mathrm{H}\mb{W}_2-2\overline{\mb{A}}\mb{A}-2\overline{\mb{B}}\mb{B}\right)\right)\nonumber\\
&&\times\exp\left(\mathrm{Tr}\left(\mb{R}_1\mb{Q}_1+\mb{R}_2\mb{Q}_2+\mb{R}_3\mb{Q}_3+\overline{\mb{R}}_4\mb{R}_4\right.\right. \nonumber\\
&&\hspace{3cm}\left.\left.+\mb{R}_5\mb{Q}_5+\mb{R}_6\mb{Q}_6-\overline{\mb{Q}}_4\mb{Q}_4+\overline{\mb{R}}_7\mb{R}_7-\overline{\mb{Q}}_7\mb{Q}_7-\mb{R}_8\mb{Q}_8\right)\right)\nonumber\\
&&\times\exp\left(-\frac{1}{2}\mathrm{Tr}\left(\frac{
\rho}{n}\mb{R}_1\mb{Y}^\mathrm{H}\mb{Y}-\mb{Q}_1\mb{W}_1^\mathrm{H}\mb{W}_2+\frac{\alpha}{n}\mb{R}_2\mb{X}^\mathrm{H}\mb{X}+\mb{Q}_2\mb{W}_2^\mathrm{H}\mb{W}_1\right)\right)\nonumber\\
&&\times \exp\left(-\frac{1}{2}\mathrm{Tr}\left(\frac{\alpha}{n}\mb{R}_3\mb{X}^\mathrm{H}\mb{X}-\mb{Q}_3\mb{Z}^\mathrm{H}\mb{W}_1+\frac{\alpha}{n}\overline{\mb{B}}\mb{W}_1\mb{R}_4+2\overline{\mb{R}}_4\mb{X}^\mathrm{H}\mb{A}\right)\right)\nonumber\\
&&\times\exp \left(-\frac{1}{2}\mathrm{Tr}\left(\frac{\alpha}{n}\mb{R}_5\mb{X}^\mathrm{H}\mb{X}-\mb{Q}_5\mb{W}_2^\mathrm{H}\mb{Z}+\frac{\alpha}{n}\mb{R}_6\mb{X}^\mathrm{H}\mb{X}+\mb{Q}_6\mb{Z}^\mathrm{H}\mb{Z}\right)\right)\nonumber\\
&&\times\exp\left(-\frac{1}{2}\mathrm{Tr}\left(-\frac{\alpha}{n}\overline{\mb{B}}\mb{Z}\mb{Q}_4-2\overline{\mb{Q}}_4\mb{X}^\mathrm{H}\mb{A}+\frac{\alpha}{n}\overline{\mb{A}}\mb{X}\mb{R}_7+2\overline{\mb{R}}_7\mb{W}_2^\mathrm{H}\mb{B}\right)\right)\nonumber\\
&&\times\exp\left(-\frac{1}{2}\mathrm{Tr}\left(-\frac{\alpha}{n}\overline{\mb{A}}\mb{X}\mb{Q}_7-2\overline{\mb{Q}}_7\mb{Z}^\mathrm{H}\mb{B}-2\frac{\alpha}{n}\mb{R}_8\overline{\mb{B}}\mb{B}-2\mb{Q}_8\overline{\mb{A}}\mb{A}
\right)\right)\nonumber\\
&&\hspace{3cm}\times d_c\mb{W}_1d_c\mb{W}_2d_c\mb{X}d_c\mb{Y}d_c\mb{Z}d_g\mb{A}d_g\mb{B}d\lambda\label{eq: moment3}\\
&=&\int\exp(-{\cal S})\cdot d\lambda\label{eq: moment4},
\end{eqnarray}
where we have combined the integral measures over the various
$\mb{R}_i$'s an $\mb{Q}_i$'s into the single integral measure
\begin{eqnarray}d\lambda&\triangleq&
d\mu(\mb{R}_1,\mb{Q}_1)d\mu(\mb{R}_2,\mb{Q}_2)d\mu(\mb{R}_3,\mb{Q}_3)d\mu(\mb{R}_5,\mb{Q}_5)\nonumber\\
&&\times d\mu(\mb{R}_6,\mb{Q}_6)d\mu(\mb{R}_8,\mb{Q}_8)\cdot
d\mb{R}_4d\mb{Q}_4 d\mb{R}_7d\mb{Q}_7.
\end{eqnarray}
In \eqref{eq: moment1} we have firstly applied \eqref{eq:
identity1} and \eqref{eq: identity2} (backwards) with $\mb{M}$ the
argument of the determinant in nominator and denominator
respectively, $\mb{N}=\mb{I}_{\nu\times\nu}$ and
$\mb{A}=\mb{B}=\mb{0}_{n\times\nu}$ in order to get rid of the
determinants. Afterwards we again applied \eqref{eq: identity1}
(backwards) twice with $\mb{M}=\mb{I}_{n\times n}$ and
$\mb{N}=\mb{I}_{\nu\times\nu}$ in order to split the products
$\mb{H}_2\mb{H}_1\mb{H}_1^\mr{H}\mb{H}_2^\mr{H}$ and
$\mb{H}_2\mb{H}_2^\mr{H}$ at the expense of the introduced
auxiliary matrices $\mb{Y}$ and $\mb{Z}$. For the first
application we have
$\mb{A}=\mb{B}=\mb{H}_1^\mr{H}\mb{H}_2^\mr{H}$, for the second one
$\mb{A}=\mb{B}=\mb{H}_2^\mr{H}$. Exactly the same is done in
\eqref{eq: moment1b} again where we also break up the products
$\mb{H}_2\mb{H}_1$ and $\mb{H}_1^\mr{H}\mb{H}_2^\mr{H}$. In
\eqref{eq: moment2} we get rid of the integrals over $\mb{H}_1$
and $\mb{H}_2$ by twice applying \eqref{eq: identity1} (forwards).
In \eqref{eq: moment3} we split all quartic terms into quadratic
terms by making use of \eqref{eq: identity2} and \eqref{eq:
identity3}. We can get rid of all integrals but the outer one, by
(forwards) applying identities \eqref{eq: identity1} and
\eqref{eq: identity2} again, and after some algebraic effort we
obtain ${\cal S}$ as
\begin{eqnarray}
{\cal S}&=&-\mathrm{Tr}\left(\mb{R}_1\mb{Q}_1+\mb{R}_2\mb{Q}_2+\mb{R}_3\mb{Q}_3+\overline{\mb{R}}_4\mb{R}_4+\mb{R}_5\mb{Q}_5\right.\nonumber\\
&&\left.+\mb{R}_6\mb{Q}_6-\overline{\mb{Q}}_4\mb{Q}_4+\overline{\mb{R}}_7\mb{R}_7-\overline{\mb{Q}}_7\mb{Q}_7-\mb{R}_8\mb{Q}_8\right)\nonumber\\
&&+n\ln\det\left(\mb{I}_{\nu}+\frac{\rho}{n}\mb{R}_1\right)+n\ln\det\left(\mb{I}_{\nu}+\frac{\alpha}{n}(\mb{R}_2+\mb{R}_3+\mb{R}_5+\mb{R}_6)\right)\nonumber\\&&+n\ln\det\left(\mb{I}_{\nu}+\mb{Q}_1\mb{Q}_2\right)+n\ln\det\left(\mb{I}_{\nu}-\left(\mb{I}_{\nu}+\mb{Q}_1\mb{Q}_2\right)^{-1}\mb{Q}_5\mb{Q}_1\mb{Q}_3+\mb{Q}_6\right)\nonumber\\
&&-n\ln\det\left(\mb{I}_{\nu}+\frac{\alpha}{n}\mb{R}_8-\frac{\alpha}{n}\left(\mb{I}_{\nu}+\mb{Q}_1\mb{Q}_2\right)^{-1}\overline{\mb{R}}_7\mb{Q}_1\mb{R}_4\right.\nonumber\\
&&\hspace{.5cm}\left.+\frac{\alpha}{n}\left[\mb{I}_{\nu}-\left(\mb{I}_{\nu}+\mb{Q}_1\mb{Q}_2\right)^{-1}\mb{Q}_5\mb{Q}_1\mb{Q}_3+\mb{Q}_6\right]^{-1}\right.\nonumber\\
&&\hspace{.75cm}\times\left[-\overline{\mb{R}}_7\mb{Q}_1\mb{Q}_3\left(\mb{I}_{\nu}+\mb{Q}_1\mb{Q}_2\right)^{-1}\mb{Q}_5\mb{Q}_1\mb{R}_4\left(\mb{I}_{\nu}+\mb{Q}_1\mb{Q}_2\right)^{-1}\right.\nonumber\\
&&\hspace{1cm}\left.\left.+\overline{\mb{R}}_7\mb{Q}_1\mb{Q}_3\mb{Q}_4\left(\mb{I}_{\nu}+\mb{Q}_1\mb{Q}_2\right)^{-1}+\overline{\mb{Q}}_7\left(\mb{I}_{\nu}+\mb{Q}_1\mb{Q}_2\right)^{-1}\mb{Q}_5\mb{Q}_1\mb{R}_4 -\overline{\mb{Q}}_7\mb{Q}_4\right]\right)\nonumber\\
&&-n\ln\det\bigg(\mb{I}_{\nu}+\mb{Q}_8+\frac{\alpha}{n}\left[\mb{I}_{\nu}+\frac{\alpha}{n}(\mb{R}_2+\mb{R}_3+\mb{R}_5+\mb{R}_6)\right]^{-1}\nonumber\\
&&\hspace{3cm}\times\left[-\overline{\mb{R}}_4\mb{R}_7+\overline{\mb{Q}}_4\mb{R}_7+\overline{\mb{R}}_4\mb{Q}_7-\overline{\mb{Q}}_4\mb{Q}_7\right]\bigg).\label{eq:
moment5}
\end{eqnarray}
At this point we have shaped the problem into the form of
\eqref{eq: explanation}, where the role of $x$ is played by the
introduced $\nu \times \nu$ auxiliary matrices. Note that there
appears no matrix with one of its dimension equal to $n$ in ${\cal
S}$ anymore.

\subsection{Evaluating Mean and Variance}
\label{sec: step2} In order to evaluate the last remaining
integral in \eqref{eq: moment4} by means of saddle point
integration we need to expand ${\cal S}$ into a Taylor series in
$\bm{\delta}\mb{ R}_1, \bm{\delta}\mb{ Q}_1,\ldots,
\bm{\delta}\mb{ R}_8, \bm{\delta}\mb{ Q}_8$ around its minimum.
This expansion corresponds to the expansion in $x$ in Section
\ref{subsec: saddlepoint}. With ${\cal S}_p$ denoting the $p$th
order term in the series the expansion looks as follows
\begin{equation}\label{eq: taylor}
{\cal S}={\cal S}_0+{\cal S}_2+{\cal S}_3+\ldots
\end{equation}
By symmetry all complex matrices are assumed to be proportional to
the identity matrix at the minimum of ${\cal S}$ (\emph{replica
symmetry}), the Grassmann matrices have to vanish in order to
obtain a real solution (by definition real numbers cannot be
Grassmann numbers, since they commute). Thus, to develop the
Taylor series \eqref{eq: taylor} in this point we write

\begin{minipage}{.48\textwidth}
\begin{eqnarray}
\mb{R}_1&=&r_1 n \mb{I}_\nu+\bm{\delta}\mb{ R}_1\\
\mb{R}_2&=&r_2 n\mb{I}_\nu+\bm{\delta}\mb{ R}_2\\
\mb{R}_3&=&r_3 n\mb{I}_\nu+\bm{\delta}\mb{ R}_3\\
\mb{R}_4&=&\bm{\delta}\mb{ R}_4\\
\overline{\mb{R}}_4&=&\bm{\delta}\mb{}\overline{\mb{
R}}_4\\
\mb{R}_5&=&r_5n \mb{I}_\nu+\bm{\delta}\mb{ R}_5\\
\mb{R}_6&=&r_6n\mb{I}_\nu+\bm{\delta}\mb{ R}_6\\
\overline{\mb{R}}_7&=&\bm{\delta}\mb{ R}_7\\
\overline{\mb{R}}_7&=&\bm{\delta}\mb{}\overline{\mb{ R}}_7\\
\mb{R}_8&=&r_8 n\mb{I}_\nu+\bm{\delta}\mb{ R}_8
\end{eqnarray}
\end{minipage}
\begin{minipage}{.48\textwidth}
\begin{eqnarray}
\mb{Q}_1&=&q_1\mb{I}_\nu+\bm{\delta}\mb{ Q}_1\\
\mb{Q}_2&=&q_2\mb{I}_\nu+\bm{\delta}\mb{ Q}_2\\
\mb{Q}_3&=&q_3\mb{I}_\nu+\bm{\delta}\mb{ Q}_3\\
\mb{Q}_4&=&\bm{\delta}\mb{ Q}_4\\
\overline{\mb{Q}}_4&=&\bm{\delta}\mb{}\overline{\mb{
Q}}_4\\
\mb{Q}_5&=&q_5\mb{I}_\nu+\bm{\delta}\mb{ Q}_5\\
\mb{Q}_6&=&q_6\mb{I}_\nu+\bm{\delta}\mb{ Q}_6\\
\overline{\mb{Q}}_7&=&\bm{\delta}\mb{ Q}_7\\
\overline{\mb{Q}}_7&=&\bm{\delta}\mb{}\overline{\mb{ Q}}_7\\
\mb{Q}_8&=& q_8\mb{I}_\nu+\bm{\delta}\mb{ Q}_8
\end{eqnarray}
\end{minipage}

By definition ${\cal S}_0$ is given by \eqref{eq: moment5}
evaluated at the minimum of ${\cal S}$, i.e.,
\begin{eqnarray}
{\cal S}_0&=&\nu \cdot n \cdot \{\ln\left(1+\rho
r_1\right)+\ln\left(1+\alpha
(r_2+r_3+r_5+r_6)\right)\nonumber\\
&&+\ln\left(1+q_1q_2-q_1q_3q_5q_2+q_6+q_1q_2q_6\right)-\ln\left(1+\alpha
r_8\right)\nonumber\\
&&-\ln\left(1+q_8\right)-(r_1q_1+r_2q_2+r_3q_3+r_5q_5+r_6q_6-r_8q_8)\}.\label{eq:
gammaraw}
\end{eqnarray}
The respective coefficients $r_i$ and $q_i$ have to ensure that
${\cal S}_1=0$. They are found by differentiating \eqref{eq:
gammaraw} for each of them and setting the resulting expressions
to zero. The derivatives for the $r_i$'s (note that we can
summarize $r_2+r_3+r_5+r_6+r_8\triangleq\tilde{r}_2$ by symmetry)
yield

\begin{minipage}{.48\textwidth}
\begin{eqnarray}
0&=&q_1-\frac{\rho}{1+\rho r_1}\\
0&=&q_2-\frac{\alpha}{1+\alpha \tilde{r}_2}\\
0&=&q_3-\frac{\alpha}{1+\alpha \tilde{r}_2}
\end{eqnarray}
\end{minipage}
\begin{minipage}{.48\textwidth}
\begin{eqnarray}
0&=&q_5-\frac{\alpha}{1+\alpha \tilde{r}_2}\\
0&=&q_6-\frac{\alpha}{1+\alpha \tilde{r}_2}\\
0&=&q_8-\frac{\alpha}{1+\alpha r_8}.
\end{eqnarray}
\end{minipage}

\vspace{.5cm} We see that $q_2=q_3=q_5=q_6$. Taking this into
account the derivatives for the $q_i$'s yield
\begin{eqnarray}
0&=&r_1-\frac{q_2}{1+q_1q_2+q_2}\\
0&=&\tilde{r}_2-\frac{1+q_1}{1+q_1q_2+q_2}\\
0&=&r_8-\frac{1}{1+ q_8}.
\end{eqnarray}
The leading term thus simplifies to
\begin{eqnarray}\label{eq: S0}
{\cal S}_0&=&\nu \cdot n \cdot \{\ln\left(1+\rho
r_1\right)+\ln\left(1+\alpha
\tilde{r}_2\right)+\ln\left(1+q_2+q_1q_2\right)\nonumber\\
&&-\ln\left(1+\alpha
r_8\right)-\ln\left(1+q_8\right)-(r_1q_1+\tilde{r}_2q_2-r_8q_8)\}\triangleq\nu\cdot
n\cdot \xi_1
\end{eqnarray}
with
\begin{center}
\begin{minipage}{.48\textwidth}
\begin{eqnarray}
q_1&=&\frac{\rho}{1+\rho r_1}\\
q_2&=&\frac{\alpha}{1+\alpha \tilde{r}_2}\\
q_8&=&\frac{\alpha}{1+\alpha r_8}
\end{eqnarray}
\end{minipage}
\begin{minipage}{.48\textwidth}
\begin{eqnarray}
r_1&=&\frac{q_2}{1+q_1q_2+q_2}\\
\tilde{r}_2&=&\frac{1+q_1}{1+q_1q_2+q_2}\\
r_8&=&\frac{1}{1+ q_8}.
\end{eqnarray}
\end{minipage}
\end{center}
We note that $\xi_1(\cdot)$ in \eqref{eq: S0} is the multivariate
version of the function mentioned in Section \ref{subsec:
saddlepoint}. We see that $\xi_1(\cdot)$ is ${\cal O}(n^0)$ and
thus $n\cdot\xi_1(\cdot)$, which will turn out to correspond to
the mean of $I$ in the large $n$ limit, is ${\cal O}(n)$.

At this point, we make use of the variable transformations
$\mb{R}_x\rightarrow \bm{\delta}\mb{ R}_x$ and
$\mb{Q}_x\rightarrow \bm{\delta}\mb{ Q}_x$ for $x=1...8$, which
preserve the integral measures. We indicate that transformation by
denoting the respective integral measure. Furthermore, we define
\begin{eqnarray}
\mb{x}_{ab}^{(1)}\!\!&\!\!\triangleq\!\!&\!\![\delta
R_{1,ab},\delta R_{2,ab},\delta R_{3,ab},\delta R_{5,ab},\delta
R_{6,ab},\delta Q_{1,ab},\delta Q_{2,ab},\delta Q_{3,ab},\delta
Q_{5,ab},\delta
Q_{6,ab}]^\mathrm{T}\\
\mb{x}_{ab}^{(2)}\!\!&\!\!\triangleq\!\!&\!\![\delta
R_{8,ab},\delta
Q_{8,ab}]^\mathrm{T}\\
\mb{x}_{ab}^{(3)}\!\!\!&\!\triangleq\!\!&\!\![\delta\overline{R}_{4,ab},\delta
R_{4,ab},\delta Q_{4,ab} ,\delta \overline{Q}_{4,ab},\delta
R_{7,ab} ,\delta \overline{R}_{7,ab},\delta Q_{7,ab} ,\delta
\overline{Q}_{7,ab}]^\mathrm{T}.
\end{eqnarray}
With this notation we can write the moment generating function in
terms of the Hessians of \eqref{eq: moment5}, $\mb{V}_1$,
$\mb{V}_2$ and $\mb{V}_3$, as defined in \eqref{eq: V1} -
\eqref{eq: V3} as
\begin{eqnarray}
g_I(\nu)&=&e^{-{\cal S}_0} \int
\exp(-{\cal S}_2-{\cal S}_3-{\cal S}_4-\ldots)\cdot d\lambda\\
&=&e^{-{\cal S}_0}\cdot \int
\exp(-{\cal S}_2)\cdot\left\{1-[{\cal S}_3+{\cal S}_4+\ldots]+\frac{1}{2}\left[{\cal S}_3+{\cal S}_4+\ldots\right]^2-\ldots\right\}\cdot d\lambda\label{eq: expand}\\
&=&e^{-{\cal S}_0} \cdot \int
\exp\left(-\frac{1}{2}\sum_{i=1}^3\sum_{a,b=1}^\nu
\mb{x}_{ab}^{(i)\mathrm{T}}\mb{V}_i\mb{x}_{ab}^{(i)}\right)\nonumber\\
&&\hspace{2cm}\times\left\{1-[{\cal S}_3+{\cal S}_4+\ldots]+\frac{1}{2}\left[{\cal S}_3+{\cal S}_4+\ldots\right]^2-\ldots\right\}\cdot d\lambda\label{eq: moment6a}\\
&=&e^{-{\cal S}_0}\cdot \Bigg[\left|\frac{\det \mb{V}_1\det
\mb{V}_2}{(\det \mb{V}_3)^{2}}\right|^{-\frac{\nu^2}{2}}+ \int
\exp\left(-\frac{1}{2}\sum_{i=1}^3\sum_{a,b=1}^\nu
\mb{x}_{ab}^{(i)\mathrm{T}}\mb{V}_i\mb{x}_{ab}^{(i)}\right)\nonumber\\
&&\hspace{2cm}\times\left\{-[{\cal S}_3+{\cal
S}_4+\ldots]+\frac{1}{2}\left[{\cal S}_3+{\cal
S}_4+\ldots\right]^2-\ldots\right\}\cdot d\lambda\Bigg].\label{eq:
moment6b}
\end{eqnarray}
In \eqref{eq: expand} we expanded $\exp(-{\cal S}_3-{\cal
S}_4-\ldots)$ into a series. The evaluation of the integral over
the first term in \eqref{eq: moment6a} is provided in
\cite{moustakas1}. We note that $\xi_2(\cdot)\triangleq-\ln\det
|\mb{V}_1|-\ln|\det \mb{V}_2|+2\ln\det |\mb{V}_3|$, which will
turn out to correspond to the variance of $I$ in the large $n$
limit, is ${\cal O}(1)$. Again, $\xi_2(\cdot)$ is the multivariate
version of the function mentioned in Section \ref{subsec:
saddlepoint}.

\subsection{Proving Gaussianity}
\label{sec: step3} We will next show, that the remaining integral
expression
\begin{equation}\label{eq: ref}
\int \exp\left(-\frac{1}{2}\sum_{i=1}^3\sum_{a,b=1}^\nu
\mb{x}_{ab}^{(i)\mathrm{T}}\mb{V}_i\mb{x}_{ab}^{(i)}\right)\cdot\left\{-[{\cal
S}_3+{\cal S}_4+\ldots]+\frac{1}{2}\left[{\cal S}_3+{\cal
S}_4+\ldots\right]^2-\ldots\right\}\cdot d\lambda
\end{equation}
is ${\cal O}(n^{-1})$. To see this we need to consider the various
Taylor coefficients of the ${\cal S}_p$ for $p>2$ first. By
inspecting \eqref{eq: moment5} we note that

\begin{enumerate}
\item[1a)] a differentiation for either $\mb{R}_1$, $\mb{R}_2$,
$\mb{R}_3$, $\mb{R}_5$, $\mb{R}_6$ or $\mb{R}_8$ yields a
multiplication by a factor $1/n$,

\item[2a)] a differentiation for either $\mb{Q}_1$, $\mb{Q}_2$,
$\mb{Q}_3$, $\mb{Q}_5$, $\mb{Q}_6$ or $\mb{Q}_8$ does not change
the order with respect to $n$,

\item[3a)] \emph{two} differentiations for Grassmann variables
(note that odd numbers of differentiations yield zero Taylor
coefficients) yield a multiplication by a factor $1/n$.
\end{enumerate}
Accordingly a Taylor coefficient resulting from $i$, $j$ and $k$
differentiations of the first, second and third type, respectively
will be ${\cal O}(n^{1-i-k/2})$. Also, a product of $t$ Taylor
coefficients resulting from $i_1$, $j_1$, $k_1$, $i_2$, $j_2$,
$k_2$, $\ldots$, $i_t$, $j_t$, $k_t$ differentiations of the
first, second and third type, each, will be ${\cal O}(n^{t-\sum_l
(i_l+k_l/2)})$.

Next, consider integrals of the form
\begin{equation}
\int \exp\left(-\frac{1}{2}\sum_{i=1}^3\sum_{a,b=1}^\nu
\mb{x}_{ab}^{(i)\mathrm{T}}\mb{V}_i\mb{x}_{ab}^{(i)}\right)
\prod_{i,i\neq 4,7} \mb{\delta R}_i\cdot\prod_{j,j\neq 4,7}
\mb{\delta Q}_j\cdot \prod_{k_1,k_2,k_3,k_4=4,7} \mb{\delta
\overline{R}}_{k_1}\mb{\delta Q}_{k_2}\mb{\delta
\overline{Q}}_{k_3}\mb{\delta R}_{k_4}\cdot d\lambda.
\end{equation}
For the complex matrices Wick's theorem allows us to split the
integral into sums of products of integrals involving only
quadratic correlations. Furthermore, it states that for odd
numbers of factors the integral evaluates to zero. Ignoring the
Grassmann matrices for the moment we can extract the order of
these correlations in the following. We define $\mb{V}$ as the
joint Hessian
\begin{equation}
\mb{V}\triangleq\left(%
\begin{array}{cc}
  \mb{V}_1 & \mb{0} \\
  \mb{0} & \mb{V}_2 \\
\end{array}%
\right)
\end{equation}
and note that $\det(\mb{V})$ is ${\cal O}(1)$. Also, we define
$\mb{x}\triangleq
[\mb{x}_{ab}^{(1),\mr{T}},\mb{x}_{ab}^{(2),\mr{T}}]^\mathrm{T}$
and denote the integral measure $d\lambda$ without all Grassmann
contributions by $d\tilde{\lambda}$. With this notation we can
extract the orders of the three kinds of arising quadratic
correlations by applying the second part of Wick's theorem:
\begin{enumerate}
\item[1b)]\begin{eqnarray} &&\int
\exp\left(-\frac{1}{2}\sum_{a,b=1}^\nu
\mb{x}^\mathrm{T}\mb{V}\mb{x}\right)\cdot\delta
R_{i,ab}\cdot\delta
R_{j,cd}\cdot d\tilde{\lambda}\nonumber\\
&&\hspace{1.5cm}=\delta_{ad}\delta_{bc}|\det(\mb{V})|^{-\frac{\nu^2}{2}}\cdot
\frac{\det(\mb{V}^{(2i-1,2j-1)})}{\det(\mb{V})}={\cal O}(n),
\end{eqnarray}
\item[2b)]\begin{eqnarray}
 &&\int \exp\left(-\frac{1}{2}\sum_{a,b=1}^\nu
\mb{x}^\mathrm{T}\mb{V}\mb{x}\right)\cdot\delta
Q_{i,ab}\cdot\delta
Q_{j,cd}\cdot d\tilde{\lambda}\nonumber\\
&&\hspace{1.5cm}=\delta_{ad}\delta_{bc}|\det(\mb{V})|^{-\frac{\nu^2}{2}}\cdot
\frac{\det(\mb{V}^{(2i,2j)})}{\det(\mb{V})}={\cal
O}(n^{-1}),\\
\end{eqnarray}
\item[3b)]\begin{eqnarray} &&\int
\exp\left(-\frac{1}{2}\sum_{a,b=1}^\nu
\mb{x}^\mathrm{T}\mb{V}\mb{x}\right)\cdot\delta
R_{i,ab}\cdot\delta
Q_{j,cd}\cdot d\tilde{\lambda}\nonumber\\
&&\hspace{1.5cm}=-\delta_{ad}\delta_{bc}|\det(\mb{V})|^{-\frac{\nu^2}{2}}\cdot
 \frac{\det(\mb{V}^{(2i-1,j)})}{\det(\mb{V})}={\cal
O}(1).
\end{eqnarray}
\end{enumerate}
By $\det(\mb{V}^{(a,b)})$ we denote the sub-determinant when the
$a$th row and the $b$th column in the matrix is deleted,
$\delta_{xy}$ denotes the Kronecker delta function. The orders
follow, since deleting odd lines/columns in $\mb{V}$ amounts to a
multiplication of the respective determinant by a factor which is
${\cal O}(n)$, while deleting even lines/columns in $\mb{V}$
amounts to a multiplication of the respective determinant by a
factor which is ${\cal O}(n^{-1})$. The Grassmannian integrations
are easily verified to yield ${\cal O}(n^0)$ factors, since also
the elements of $\mb{V}_3$ are ${\cal O}(n^0)$.

Combining 1a) and 1b), 2a) and 2b) as well as 3a) and 3b), we can
finally summarize, that terms resulting from the evaluation of
\eqref{eq: ref} are
\begin{equation} \nonumber
{\cal O}\left(n^{t-\sum_{x=1}^t
\frac{i_x+j_x+k_x}{2}}\right), \textrm{ if }\sum_{x=1}^t
i_x+j_x+k_x \textrm { is even}
\end{equation}
or zero otherwise. Here, $t$ denotes the number of involved Taylor
coefficients, $i_x, j_x, k_x$ the number of derivations of kind 1,
2 and 3. Note that $i_x, j_x$ and $k_x$ also correspond to the
number of factors arising with the Taylor coefficient in the
correlation. Since $\sum_{x=1}^t \frac{i_x+j_x+k_x}{2}>t$ for
$p>2$, we conclude that all appearing terms in the integral are
${\cal O}(n^{-1})$ or smaller.

We can thus rewrite \eqref{eq: moment6b} as
\begin{equation}
g_I(\nu=)e^{-{\cal S}_0}\cdot \Bigg[\left|\frac{\det \mb{V}_1\det
\mb{V}_2}{(\det \mb{V}_3)^{2}}\right|^{-\frac{\nu^2}{2}}+ {\cal
O}(n^{-1})\Bigg].
\end{equation}
After factoring out the determinant the cumulant generating
function is given by
\begin{eqnarray}
\ln g_I(\nu)&=&\ln\left\{ e^{-{\cal S}_0}\left|\frac{\det
\mb{V}_1\det \mb{V}_2}{(\det
\mb{V}_3)^{2}}\right|^{-\frac{\nu^2}{2}}\left(1+\left|\frac{\det
\mb{V}_1\det \mb{V}_2}{(\det
\mb{V}_3)^{2}}\right|^{\frac{\nu^2}{2}} \cdot{\cal
O}(n^{-1})\right)\right\}\\
&=&-\nu \cdot n \cdot \xi_1-\frac{\nu^2}{2}(\ln|\det(\mb{V}_1)|+\ln|\det(\mb{V}_2)|\nonumber\\
&&\hspace{3cm}-2\ln|\det(\mb{V}_3)|)+\ln(1+{\cal
O}(n^{-1}))\\
&=&-\nu \cdot n \cdot \xi_1+\frac{\nu^2}{2}\cdot\xi_2 +{\cal
O}(n^{-1})\label{eq: cumulant}.
\end{eqnarray}

A coefficient comparison with \eqref{eq: cumgen} immediately
reveals
\begin{equation}
\mathsf{E}[I]=n\xi_1+{\cal O}(n^{-1}),
\end{equation}
and
\begin{equation}
\mathsf{Var}[I]=\xi_2+{\cal O}(n^{-1}).
\end{equation}
Also the ${\cal C}_p$ for $p>2$ are ${\cal O}(n^{-1})$ and thus
vanish for $n\rightarrow \infty$. This implies that $I$ is
Gaussian distributed in this limit. Note, that indeed the residual
term of the variance can be shown to be ${\cal O}(n^{-2})$ in the
same way as it is done in \cite{moustakas1}. The reason behind
this is that no ${\cal O}(n^{-1})$ term proportional to $\nu^2$ is
generated in \eqref{eq: moment6b}. We skip this (in the present
case very tedious) derivation for reasons of brevity.

\subsection{Reintroducing Covariance Matrices}
Finally, we reintroduce the omitted covariance matrices
$\mb{T}_\mr{s}$, $\mb{R}_\mr{r}$, $\mb{T}_\mr{r}$,
$\mb{R}_\mr{d}$. In \eqref{eq: moment1b} we see that the
covariance matrices could be attached to the introduced auxiliary
matrices as follows: $\mb{Y}^\mathrm{H}\mb{T}_\mr{s}^\frac{1}{2}$,
$\mb{R}_\mr{r}^\frac{1}{2}\mb{W}_2$,
$\mb{W}_2^\mathrm{H}\mb{T}_\mr{r}^\frac{1}{2}$,
$\mb{R}_\mr{d}^\frac{1}{2}\mb{X}$,
$\mb{X}^\mathrm{H}\mb{R}_\mr{d}^\frac{1}{2}$,
$\mb{T}_\mr{r}^\frac{1}{2}\mb{W}_1$
,$\mb{W}_1^\mathrm{H}\mb{R}_\mr{r}$,
$\mb{T}_\mr{s}^\frac{1}{2}\mb{Y}$,
$\mb{Z}^\mathrm{H}\mb{T}_\mr{r}^\frac{1}{2}$,
$\mb{R}_\mr{d}^\frac{1}{2}\mb{X}$,
$\mb{X}^\mathrm{H}\mb{R}_\mr{d}^\frac{1}{2}$,
$\mb{T}_\mr{r}^\frac{1}{2}\mb{Z}$,
$\overline{\mb{B}}\mb{T}_\mr{r}^\frac{1}{2}$,
$\mb{R}_\mr{d}^\frac{1}{2}\mb{A}$,
$\overline{\mb{A}}\mb{R}_\mr{d}^\frac{1}{2}$ and
$\mb{T}_\mr{r}^\frac{1}{2}\mb{B}$. In \eqref{eq: moment3}, we
always obtain products involving only identical (square roots of)
covariance matrices as factors. Thus, we can attach a factor
$\mb{T}_\mr{s}$ to $\mb{R}_1$, a factor $\mb{R}_\mr{r}$ to
$\mb{Q}_1$, factors $\mb{T}_\mr{r}$ to $\mb{Q}_2$, $\mb{Q}_3$,
$\mb{Q}_5$, $\mb{Q}_6$, $\mb{R}_8$, $\mb{R}_4$, $\mb{Q}_4$,
$\overline{\mb{R}}_7$ and $\overline{\mb{Q}}_7$, and factors
$\mb{R}_\mr{d}$ to $\mb{R}_2$, $\mb{R}_3$, $\mb{R}_5$, $\mb{R}_6$,
$\mb{Q}_8$, $\overline{\mb{R}}_4$, $\overline{\mb{Q}}_4$,
$\mb{R}_7$, and $\mb{Q}_7$. In \eqref{eq: moment5} these factors
are combined in outer products, while the factor of $n$ is removed
and the $\mb{I}_\nu$ are replaced by $\mb{I}_{\nu\cdot n}$. It is
then obvious, that \eqref{eq: S0} translates to \eqref{eq: mean},
and also the entries of the Hessians \eqref{eq: V1} - \eqref{eq:
V3} follow immediately. From the dimension of the covariance
matrices we can now also conclude the respective antenna array
dimension and thus also replace the $n$ by either $n_\mr{d}$,
$n_\mr{r}$ or $n_\mr{d}$ again.


\section{Comparison with Simulation Results}
We verify the results stated in the theorem by means of computer
experiments. For the mean this is done through Monte Carlo
simulations. The respective plot is shown in Fig. \ref{fig:
I_vs_snr}, where we present the ergodic mutual information versus
the SNR for $n=n_\mr{s}=n_\mr{r}=n_\mr{d}=2,4$ and $8$. We observe
that even for only two antennas the approximation is reasonable,
for four antennas the match is close to perfect, while for eight
antennas no difference between analytic approximation and numeric
evaluation can be seen anymore. In order to also verify our
results for the higher cumulant moments we compare the empirical
cumulative distribution function (CDF) of the mutual information
to a Gaussian CDF with mean and variance given in the theorem. The
respective plot is shown in Fig. \ref{fig: cdf}. Again, we observe
that the analytic approximation becomes tight indeed as
$n=n_\mr{s}=n_\mr{r}=n_\mr{d}$ increases. For $n=8$ even the tails
of the distribution are reasonably approximated, which is an
important issue for the characterization of the outage capacity.
Our simulation results thus also demonstrate that the replica
method -- despite its deficiency of not being mathematically
rigorous yet -- indeed reveals the correct solution to our
problem.

\section{Conclusion}
\label{sec:conclusion} Using the framework developped in
\cite{framework} and \cite{moustakas1} we evaluated the cumulant
moments of the mutual information for MIMO amplify and forward
relay channels in the asymptotic regime of large antenna numbers.
Similarly to the case of ordinary point-to-point MIMO channels, we
observe that all cumulant moments of order larger than two vanish
as the antenna array sizes grow large and conclude that the
respective mutual information is Gaussian distributed. For mean
and variance we obtain expressions that allow for an analytic
evaluation. Computer experiments show, that the derived
expressions serve as excellent approximations even for channels
with only very few antennas. The results confirm the linear
scaling of the ergodic mutual information (${\cal O}(n)$) in the
antenna array size and also reveal that the respective variance is
${\cal O}(1)$ in the antenna number.

\section{Acknowledgement}
The authors would like to thank Aris Moustakas for various very
valuable advices.

\appendices
\section{Preliminaries of Grassmann Variables} \label{sec:
grassmann} Grassmann algebra is a concept from mathematical
physics. A Grassmann variable (also called an anticommuting
number) is a quantity that anticommutes with other Grassmann
numbers but commutes with (ordinary) complex numbers. With
$\theta_1,\theta_2$ Grassmann variables and $\lambda$ a complex
number the defining properties are
\begin{eqnarray}
\lambda \theta_1&=&\theta_1\lambda\\
\theta_1\theta_2&=&-\theta_2\theta_1.
\end{eqnarray}
With $\theta_3$ another Grassmann variable further properties are
\begin{eqnarray}
\theta_1(\theta_2\theta_3)&=&\theta_3(\theta_1\theta_2)\\
\theta_1^2&=&0\\
\exp(\theta_1\theta_2)&=&1+\theta_1\theta_2.
\end{eqnarray}

Integration over Grassmann variables is defined by the following
to properties
\begin{eqnarray}
\int d\theta &=&0\\
\int \theta d\theta&=&1.
\end{eqnarray}
Note that also the differentials are anticommuting, i.e.,
$d\theta_1d\theta_2=-d\theta_2d\theta_1$. Further details about
integrals over Grassmann variables such as variable transformation
can be found in the Appendix of \cite{moustakas1}.

\section{Wick's Theorem}
With $\mb{V}\in \mathbb{C}^{N\times N}$, $\mb{x}\in
\mathbb{C}^{N\times 1}$ and an integral measure
$d\alpha(\mb{x})=1/\sqrt{2\pi}dx_1,\ldots,dx_N$ we have
\begin{eqnarray}\label{eq: wick}
&&(\det V)^\frac{1}{2} \int \exp\left(-\frac{1}{2}
\mb{x}^\mathrm{T}\mb{V}\mb{x}\right)\cdot\prod_{k=1}^M x_k \cdot
d\alpha(\mb{x})\\
&&\hspace{1cm}=\sum_{\mr{pairs}} (\det V)^\frac{1}{2} \int
\exp\left(-\frac{1}{2} \mb{x}^\mathrm{T}\mb{V}\mb{x}\right)\cdot
x_{i,1}\cdot x_{i,2} \cdot d\alpha(\mb{x})\cdot \\
&&\hspace{2cm}\times\ldots\\
&&\hspace{2cm} \times (\det V)^\frac{1}{2} \int
\exp\left(-\frac{1}{2} \mb{x}^\mathrm{T}\mb{V}\mb{x}\right)\cdot
x_{i_{M-1}}\cdot x_{i_M} \cdot d\alpha(\mb{x})
\end{eqnarray}
if $M$ is even. For odd $M$ the expression evaluates to zero. The
sum in \eqref{eq: wick} is over all possible rearrangements of the
orderings of the indexes such that different indexes are paired
with each other (with each distinct pairing being counted once).

Furthermore,  we have that
\begin{equation}
(\det V)^\frac{1}{2} \int \exp\left(-\frac{1}{2}
\mb{x}^\mathrm{T}\mb{V}\mb{x}\right)\cdot x_i x_j \cdot
d\alpha(\mb{x})=[V^{-1}]_{i,j},
\end{equation}
with $[V^{-1}]_{i,j}$ the element in the $i$th row and $j$th
column of $V^{-1}$. We will also need that
\begin{equation}
[V^{-1}]_{i,j}=\det \mb{V}^{(i,j)}/\det\mb{V}
\end{equation}
with $\mb{V}^{(i,j)}$ an $N-1\times N-1$ matrix, where the $i$th
row and the $j$th column of $V^{-1}$ are deleted.

\begin{figure}
\centering
\includegraphics[width=\textwidth]{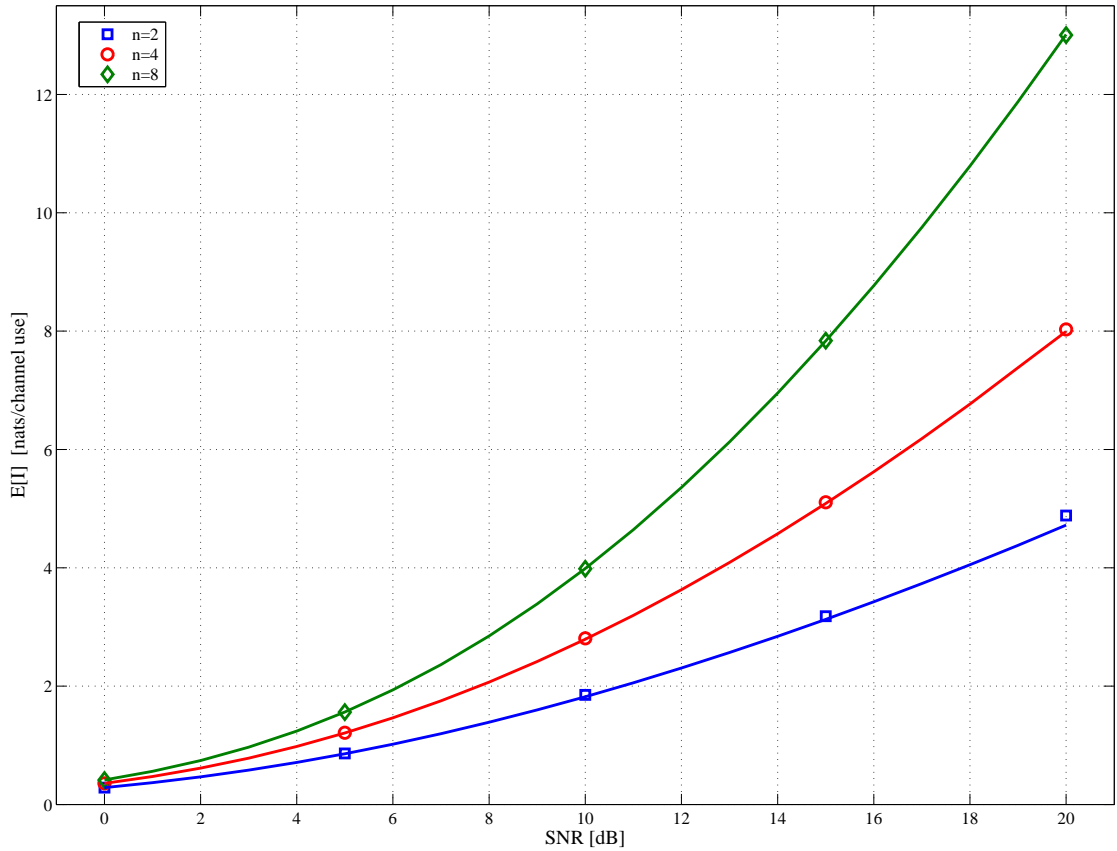}
\caption{Mutual information versus SNR for
$n=n_\mr{s}=n_\mr{r}=n_\mr{d}$ and i.i.d. channel matrix entries
-- solid lines are analytical approximations, circles, squares and
diamonds mark true mutual information as obtained through Monte
Carlo simulations.}\label{fig: I_vs_snr}
\end{figure}

\begin{figure}
\centering
\includegraphics[width=\textwidth]{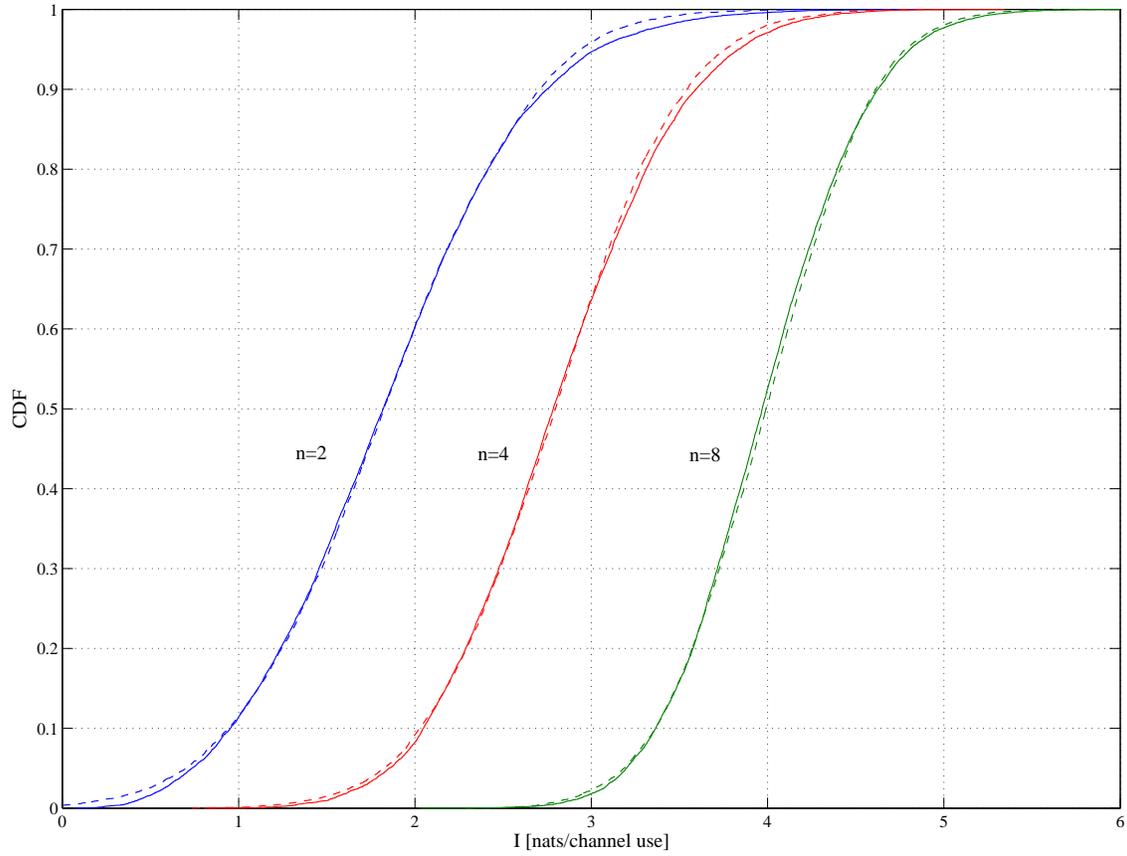}
\caption{Cumulative distribution function of mutual information
for $n=n_\mr{s}=n_\mr{r}=n_\mr{d}$ and i.i.d. channel matrix
entries. Dashed lines represent Gaussian distributions with
analytically computed mean and variance. The solid lines are the
empirical distributions obtained through simulations.\label{fig:
cdf}}
\end{figure}

\bibliography{refs}
\bibliographystyle{ieeetran}

\newpage
%

\end{document}